\documentclass[pdflatex,sn-mathphys-num]{sn-jnl}

\usepackage{amsfonts,amssymb}
\usepackage{amsmath}
\usepackage{graphicx}
\usepackage{epstopdf}
\usepackage{enumitem}
\usepackage{caption,subcaption}

\usepackage{comment}

\setlist[enumerate]{leftmargin=.5in}
\setlist[itemize]{leftmargin=.5in}

\usepackage{algpseudocode}

\usepackage{dsfont}

\usepackage{color}

\newcommand{\one}{{\mathbf{1}}}
\newcommand{\est}{{\hfill $\star$}}

\newcommand{\tran}{^{\top}}

\newcommand{\diag}{\mbox {\rm diag}}

\newcommand{\beq}{\begin{equation}}
\newcommand{\eeq}{\end{equation}}
\newcommand{\bea}{\begin{eqnarray}}
\newcommand{\eea}{\end{eqnarray}}
\newcommand{\beas}{\begin{eqnarray*}}
\newcommand{\eeas}{\end{eqnarray*}}
\newcommand{\ba}{\begin{array}}
\newcommand{\ea}{\end{array}}
\newcommand{\bit}{\begin{itemize}}
\newcommand{\eit}{\end{itemize}}
\newcommand{\ben}{\begin{enumerate}}
\newcommand{\een}{\end{enumerate}}

\newcommand{\ped}[1]{ _{ {\mathrm{#1} } }}

\newcommand{\ap}[1]{ ^{ {\mathrm{#1} } }}

\newcommand{\Real}[1]{ { {\mathbb R}^{#1} } }

\newcommand{\calC}{{\cal C}}

\newcommand{\calL}{{\cal L}}

\newcommand{\calI}{{\cal I}}
\newcommand{\calE}{{\cal E}}

\newcommand{\calV}{{\cal V}}

\newcommand{\lam}{\lambda}

\definecolor{lgray}{gray}{0.7}

\newtheorem{remark}{Remark}
\newtheorem{definition}{Definition}
\newtheorem{proposition}{Proposition}
\newtheorem{assumption}{Assumption}
\newtheorem{thm}{Theorem}

\begin{document}

\title[Default Robustness and Worst-Case Losses in Financial Networks]{Default Robustness and Worst-Case Losses in Financial Networks}

% Authors: full names plus addresses.
\author*{\fnm{Giuseppe C.} \sur{Calafiore}}\email{giuseppe.calafiore@polito.it}

\author{\fnm{Giulia} \sur{Fracastoro}}\email{giulia.fracastoro@polito.it}
\author{\fnm{Anton V.} \sur{Proskurnikov}}\email{anton.p.1982@ieee.org}
\equalcont{Authors contributed equally to this work.}

\affil{\orgdiv{DET} - \orgname{Politecnico di Torino},
\orgaddress{\street{Corso Duca Degli Abruzzi 24}, \city{Torino}, \country{Italy}}}
%city={Torino},
%postcode={10129},
%country={Italy}}

\abstract{We analyze the resilience and worst-case losses in a financial network of banks linked by mutual liabilities and shared exposures to external assets. Abrupt asset price changes lead to simultaneous shocks to bank balance sheets, potentially triggering cascades of defaults.
In this context, we introduce first the concept of {\em default resilience margin}, $\epsilon^*$, defined as the maximum amplitude of asset prices fluctuations that the network can sustain without generating defaults.
Such threshold value is computed by considering two different measures of price fluctuations, one based on the maximum individual variation of each asset, and the other based on the sum of all the asset's absolute variations. For any price perturbation having amplitude no larger than $\epsilon^*$, the network absorbs the shocks and remains free of  defaults. When the perturbation amplitude goes {\em beyond} the $\epsilon^*$ level, however, defaults may occur.
We assume that external liabilities have higher priority than interbank claims, thereby distinguishing between banks that default on obligations to other banks and those that are fully insolvent, i.e., unable to meet even their external obligations. We propose an explicit method to determine the upper bound on shock amplitude, $\epsilon\ped{ub}$ below which such insolvencies do not occur -- that is, all banks are able to fulfill their external liabilities, though some may default on their interbank obligations. For each shock amplitude $\epsilon\in (\epsilon^*,\epsilon\ped{ub})$,
we show how to compute the worst-case systemic loss, that is, the total financial network shortfall under the worst-case scenario of price variation of given magnitude. %Both the threshold level $\epsilon^*$ and the worst-case loss can be efficiently determined using linear programming.
}


\maketitle

\section{Introduction}

The global financial crisis of 2007-2008 sparked a surge of research into the vulnerabilities of financial systems, their susceptibility to shocks, and the phenomenon of financial contagion~\cite{Glasserman2016,Pacelli2025,Allen2000}. Significant attention has been dedicated to unraveling the intricate relationships between financial institutions and capital flows, with  a particular emphasis on their impact on the resilience of the global financial system -- its ability to absorb and insulate itself from external shocks. This body of research underlines the critical role of interconnectedness in determining stability of financial systems.

\paragraph{Resilience and Fragility in Interbank Networks.} The link between a financial network's structure and its resilience to shock propagation is complex and not yet fully understood. On one hand, dense interconnections can help redistribute liquidity and provide implicit insurance against moderate shocks. Through bilateral agreements, banks effectively share portfolio losses across many counterparties, making the system more robust to moderate external disturbances, as shown in models such as~\cite{Allen2000,Freixas2000,Babus2016}. On the other hand, dense networks are more vulnerable to severe shocks that surpass critical thresholds or affect key nodes. Interconnections facilitate shock transmission, increasing the risk of cascading defaults~\cite{Glasserman2016}. This vulnerability has been proven in regular networks~\cite{Acemoglu2015} and supported by simulations on random graphs~\cite{Nier2007,Hurd2023,Erol2023}, which show that greater connectivity can either enhance or undermine resilience depending on the nature and scale of the shock. Both effects can be amplified by the multiplex structure of financial networks, where different layers represent distinct types of interbank dependencies and assets~\cite{delRioChanona2020,Jurakovaite2024,Zino2025}.

Beyond the transmission of payment shortfalls, financial contagion can also stem from asset commonality—banks' shared exposures to declines in the value of external assets~\cite{Cifuentes2005,Glasserman2016,Allen2012}. Such common exposures can trigger systemic liquidity shocks that affect all holders of the impacted assets~\cite{Banerjee2022}, regardless of direct interbank connections. Nevertheless, network structure can amplify these risks in several ways. An initial asset shock may lead to defaults, prompting distressed banks to conduct ``fire sales'' of their holdings~\cite{Detering2021}. These sales depress asset prices further, potentially igniting information contagion~\cite{Glasserman2016}, where falling confidence and fear of widespread defaults drive down the value of even sound assets, affecting the whole financial system. Circular mechanisms of financial contagion -- particularly those driven by fire sales at distressed banks -- have been examined in several studies (see, e.g.,~\cite{BarMahSil:2021} and the recent overview in~\cite{CaiazzoZazzaro:2025}, which also provides pointers to related literature).

\paragraph{Systemic Loss Quantification under External Shocks.}
The Eisenberg–Noe framework~\cite{Eisenberg2001} is one of the most widely recognized models for analyzing financial contagion and evaluating the impact of default cascades in interbank networks. Their model addresses the clearing of mutual liabilities in the event of institutional defaults triggered by external shocks, offering an explicit method to compute both the clearing payment matrix and the total shortfall, or systemic loss~\cite{Glasserman2016}. Extensions of the Eisenberg–Noe model have incorporated key features of real-world financial systems, including the prioritization of liabilities to the non-financial sector over interbank claims~\cite{Elsinger2006}, default costs~\cite{Rogers2013}, multiple debt maturities~\cite{Kusnetsov2019}, asset liquidation costs and strategies~\cite{Feinstein2017,Feinstein2023}, cross-holdings~\cite{Weber2017}. The computation and uniqueness of the clearing payment under external shocks have been extensively studied in the literature~\cite{Massai2022,Csoka2024,Kanellopolous2024}. Additional lines of research have addressed more specific issues, such as the sensitivity of the clearing payments to small liquidity shocks~\cite{Liu2010} and to slight changes in interbank liabilities~\cite{Feinstein2018}. The Eisenberg-Noe framework has also served as a foundation for the development of dynamic models of clearing~\cite{Calafiore2023,Banerjee2025} with time-evolving obligations and payments. Most of the aforementioned research, however, assumes that the asset and liability positions of all banks are fully known.

Since banks are typically exposed to common assets, fluctuations in market prices generate shocks that are, on the one hand, difficult to predict and, on the other hand, impact the balance sheets of all affected institutions. Existing research that accounts for these uncertainties primarily focuses on random asset values, as reflected in the extensive literature on systemic risk measures (see~\cite{AraratMeimanjan2023,Jarrow2024} and references therein), network valuation theory~\cite{Barucca2020,Pallante2025}, and topology optimization to enhance resilience to random shocks~\cite{LiZhang2024} or to identify network configurations most prone to contagion~\cite{hu2024robust}.

\paragraph{The Problem in Question and Contributions.} In this paper, we address a novel class of problems in systemic risk evaluation under asset commonality. Unlike existing approaches, asset price fluctuations are not assumed to follow known probabilistic distributions but are instead constrained in magnitude -- for instance, each asset price may vary by at most~$\varepsilon$. Within this framework, we explore two key questions: (1) What is the maximum level of asset price variation that a banking system can withstand without triggering any defaults? and (2) once this resilience threshold is exceeded, what is the worst-case systemic loss resulting from cascaded defaults within the financial network? Our analysis builds on an extended version of the Eisenberg-Noe model, introduced in~\cite{Elsinger2006}, which permits banks to experience a negative net liquidity inflow from the external sector -- potentially leading to insolvency. In addition, we assume that the asset side of some banks’ balance sheets includes, beyond interbank claims and cash flows from the non-financial sector, stakes in a set of external assets.

We examine two distinct categories of uncertain fluctuations in external asset prices: (i) a scenario in which the total sum of absolute variations in asset prices does not exceed a predetermined threshold of $\epsilon$ units (referred to as an $\ell_1$ price perturbation), and (ii) a scenario in which the price of each individual asset fluctuates by no more than $\epsilon$ units (referred to as an $\ell_{\infty}$ price perturbation). We evaluate the network's resilience against the most severe shock of each type and assess the worst-case systemic loss caused by these shocks.
%The computational results we propose are \emph{exact} and can be computed efficiently using linear programming.
%
The contributions of this paper are as follows:
\begin{enumerate}
\item We compute the network's \emph{default resilience margin}, denoted by $\epsilon^*$, which represents the largest value of $\epsilon$ such that the entire system remains default-free under all asset price perturbations with amplitude not exceeding~$\epsilon$.
    %Within this range, the network can fully absorb shocks through liquidity redistribution and reductions in net worth at specific nodes.
    The computation of $\epsilon^*$ %, as detailed in Section~\ref{sec:primedefaults},
    provides a quantitative measure of the financial network's \emph{robustness} to asset price fluctuations.
\item When the price perturbation level $\epsilon$ exceeds the threshold $\epsilon^*$, defaults may occur at some nodes, leading to the problem of evaluating the \emph{worst-case systemic loss} (the total shortfall of financial institutions). The term ``worst-case'' refers to the maximum systemic loss resulting from the most adverse shock of magnitude not exceeding $\epsilon$.
It is important to note that the loss function is highly nonlinear in $\epsilon$ and lacks a closed-form analytical expression, necessitating numerical methods. By leveraging linear programming duality, we show that for values of $\epsilon$ exceeding $\epsilon^*$ but bounded above by a certain threshold $\epsilon_{\mathrm{ub}}$, both the worst-case loss and the corresponding worst-case shock can be efficiently computed by solving a linear programming problem.
\item We show that $\epsilon_{\mathrm{ub}} \geq \epsilon^*$ represents the largest shock magnitude the network can withstand without any bank becoming fully insolvent -- that is, unable to meet its external liabilities even with all interbank payments suspended. Once $\epsilon$ exceeds this threshold, there exist shocks of magnitude $\epsilon$ that lead to insolvency at one or more banks. In contrast, for all $\epsilon \leq \epsilon_{\mathrm{ub}}$, every bank remains solvent with respect to its external obligations. We compute the value $\epsilon_{\mathrm{ub}} \geq \epsilon^*$, henceforth called the \emph{insolvency resilience margin}.
\end{enumerate}

\paragraph{Organization of the paper.}
The remainder of the paper is structured as follows.
Section~\ref{sec:fin_nets} presents a method for evaluating systemic loss caused by external shocks, based on a generalized Eisenberg–Noe model and the concept of a clearing vector.
The main contributions are summarized in Section~\ref{sec:main}. In particular, Subsection~\ref{sec:primedefaults} addresses the computation of the default resilience margin; Subsection~\ref{sec:worst-case} focuses on the analysis and computation of the worst-case systemic loss; and Subsection~\ref{sec:upperlimit} presents the computation of the insolvency resilience threshold. Our principal findings are illustrated via numerical examples in Section~\ref{sec:numeric}. Conclusions are drawn in Section~\ref{sec:conclusion}.
 For improved readability, the technical proofs and other details of our key results are contained in the Appendix.

\section{Models and Methods}\label{sec:fin_nets}

\textbf{Notation.}
%\subsection{Preliminaries and notation}\label{sec:notation}
%
Given a finite set $\calV$, the symbol $|\calV|$ stands for its cardinality. For two vectors $a,b\in\mathbb{R}^n$, we write $a\leq b$ if $a_i\leq b_i$ $\forall i$.
The relation $\geq$ for vectors is defined similarly. If $a\leq b$, then $[a,b]$ denotes the set of vectors $c$ such that $a\leq c\leq b$. We write $a\geq 0$ if $a_i\geq 0\,\forall i$. Operations $\min,\max$ are applied to vectors element-wise, e.g., $\max(a,b)=(\max(a_i,b_i))_{i=1}^n$.
Given a vector $a$, we denote its positive part by $[a]^+\doteq\max(a,0)$. For a real number $z\in\mathbb{R}$, let
 $\mbox{sign}(z) \doteq 1$ if $z>0$, $\mbox{sign}(z) \doteq -1$ if $z<0$ and
 $\mbox{sign}(z) \doteq 0$ if $z=0$. The symbol $\one$ represents a (column) vector of all ones, with dimension inferred from context.

\subsection{Financial Networks with Asset Commonality}

We follow the basic model introduced in~\cite{Eisenberg2001,Glasserman2016}, with the added assumption that banks are also interconnected through asset commonality~\cite{Feinstein2023}. The parameters of the financial network, described in the next paragraphs, are summarized in Table~\ref{tab:parameters}.

\textbf{Interbank liabilities.} A \emph{financial network} as a weighted directed graph ${\cal G}=({\calV},{\calE},{\bar P})$, where the node set $\calV \doteq \{1, 2, \ldots, n\}$ represents financial institutions (e.g., banks, funds, insurance companies). The weighted adjacency matrix $\bar P = (\bar p_{ij})$ encodes nominal mutual liabilities, with $\bar p_{ij} \geq 0$ indicating that institution $i$ owes $\bar p_{ij}$ currency units to institution $j \ne i$.
The liability arises when institution $i$ borrows funds from institution $j$ under a contractual agreement -- such as an interbank loan, repurchase agreement, or other credit arrangement. We assume that $\bar p_{ii} = 0$ for all $i\in\calV$. An arc $(i, j) \in \calE$ exists if and only if $\bar p_{ij} > 0$, indicating that node $i$ has a liability to node $j$.

The total liability of node~$i$ to other nodes is given by its weighted out-degree, $\bar p_i \doteq \sum_{j \ne i} \bar p_{ij}$. It is convenient to define the vector of total liabilities as $\bar p \doteq (\bar p_i)_{i \in \calV} = \bar P \one$.

\textbf{Assets and Net Liquidity Positions.} In addition to interbank obligations, banks are interconnected through portfolio overlaps~\cite{Feinstein2023}, reflecting shared exposures to common external assets (e.g., government and corporate bonds, publicly traded stocks, commodities, and related financial instruments).

Specifically, we assume that banks may hold shares in a set of marketable assets, denoted by $\mathcal{M} \doteq \{1, \ldots, m\}$, with corresponding prices $v_1, \ldots, v_m\geq 0$. Bank $i$ holds $s_{ij}$ shares\footnote{Although $s_{ij} \geq 0$ in standard cases—representing a \emph{long} position of bank~$i$ in asset~$j$ -- the methodology developed here also accommodates \emph{short} positions, allowing some entries $s_{ij}$ to be negative. A negative value of $s_{ij}$ indicates that bank~$i$ has borrowed and sold $|s_{ij}|$ shares of asset~$j$, anticipating a price decline and intending to repurchase them later at a lower price.} of asset $j$, and the total market value of its external asset portfolio is given by $z_i \doteq \sum_{j=1}^m s_{ij} v_j = \sigma_i^\top v$, where $\sigma_i^\top \doteq [s_{i1}, \ldots, s_{im}]$ and $v^\top \doteq [v_1, \ldots, v_m]$.

Each bank has a net liquidity position $c_i^e \in \mathbb{R}$ from the external sector, representing the difference between incoming cash flows and outgoing funding obligations to entities outside the financial network. The total external net position of bank~$i$ -- that is, the total amount it receives from outside the financial system -- is thus given by $c_i \doteq c_i^e + z_i$, for all $i \in \calV$.
Denoting with $S=(s_{jk})$ the $n\times m$ matrix of asset shares and with $v$ the $m$-vector of asset prices, the vector $z$
containing the total asset values held by the banks is given by $z\doteq S v$, and the net inflow vector is thus found as
\beq
c = c^e + Sv.
\label{eq:extassets}
\eeq
It should be noted that the net external position $c_i$ can be negative\footnote{As discussed in~\cite{Elsinger2006}, the condition $c_i \geq 0$ can only be guaranteed when liabilities to the external sector are treated as having equal seniority with interbank liabilities—a simplifying assumption adopted by Eisenberg and Noe in~\cite{Eisenberg2001}. However, if external liabilities are given priority over interbank claims, it is possible for a bank’s obligations to the external sector to exceed the total value of its assets.}. We represent it as $c_i = c_i^+ - c_i^-$, where $c_i^+ \geq 0$ denotes the total external inflow to bank~$i$ (including both cash inflows and the market value of held assets), and $c_i^- \geq 0$ represents its liabilities to the external sector.
With this notation, the nominal asset and liability sides of the balance sheet for bank~$i$ are given, respectively, by
\begin{equation}\label{eq.in-out-nominal}
\bar\phi_i^{\mathrm{in}}(c) \doteq c_i^+ + \sum\nolimits_{k \ne i} \bar p_{ki}, \quad
\bar\phi_i^{\mathrm{out}}(c) \doteq c_i^- + \bar p_i.
\end{equation}
Here, $\bar\phi_i^{\mathrm{in}}(c)$ represents the total nominal assets of bank~$i$—consisting of external inflows and incoming interbank payments—while $\bar\phi_i^{\mathrm{out}}(c)$ represents its total nominal liabilities to both external and interbank creditors.

\begin{table}[tb]\caption{Parameters of a financial network}
    \label{tab:parameters}
    \centering
\begin{tabular}{r|l}
$\mathcal V=\{1,\ldots,n\}$&set of banks\\
$\mathcal M=\{1,\ldots,m\}$&set of external assets\\
$\bar p_{ij}$&liability from bank $i\in\mathcal V$ to bank $j\in\mathcal V$\\
$\bar p_{i}$&total liability of bank $i\in\mathcal V$\\
$p_{ij}$&actual clearing payment bank $i\in\mathcal V$ to bank $j\in\mathcal V$\\
$ c_{i}^e$&net liquidity  of bank $i\in\mathcal V$\\
$s_{ij}$&shares of asset $j\in\mathcal M$ owned by bank $i\in\mathcal V$\\
$v_{j}$&value per shares of asset $j\in\mathcal M$\\
$c_i$&external net position of bank $i\in\mathcal V$\\
%$z_{i}$&value of the external assets held by bank $i\in\mathcal V$\\
    \end{tabular}
\end{table}

%\begin{remark}\rm
%Although $s_{ij} \geq 0$ in standard situations, corresponding to bank $i$ having a \emph{long} position in asset $j$, the methodology developed here can also handle \emph{short}  positions, where some entries $s_{ij}$ can be negative.\erem
%\end{remark}
%\begin{remark}\label{rem.2}\rm
%Building upon the framework introduced in~\cite{Elsinger2006}, we allow banks to have negative net inflows, thereby relaxing the key assumption of the original Eisenberg-Noe model that $c_i\geq 0$. As discussed in~\cite{Elsinger2006,Hurd2016}, the restriction $c_i\geq 0$ is realistic only if the liabilities to the external sector have the same priority as the interbank liability. If the external liabilities take precedence over the interbank debt claims, as we assume in this paper, then we allow
%$c_i$ to be negative and we
%represent it as
%$c_i=c_i^+-c_i^-$, where the vector $c_i^+\geq 0$ represents the external inflow of bank $i$ (including the cash inflow and total value of assets), and $c_i^-\geq 0$ %represents its liability to the external sector that needs to be cleared before the interbank clearing. This net inflow $c_i$ can be positive or negative.
%\erem
%\end{remark}

\subsection{Actual payments, clearing vectors and system loss}

Under regular operating conditions, the nominal external net value vector $c = \bar c$ is such that $\bar\phi_i^{\mathrm{in}}(\bar c) \geq \bar\phi_i^{\mathrm{out}}(\bar c)$ for all $i$. This indicates that each bank is able to meet its obligations at the end of the period, possibly by liquidating some of its assets.

However, if a financial shock affects certain nodes -- such as a decline in the value of external assets below expected levels—then $c_j < \bar c_j$ for some $j$, and it may result in an institution~$i$ being unable to fully meet its payment obligations, as its liabilities exceed its assets, that is,
$\bar\phi_i^{\mathrm{out}}(\bar c) > \bar\phi_i^{\mathrm{in}}(\bar c)$.
In this situation of \emph{default}, node~$i$ pays out according to its available resources: it first satisfies its obligations to the external sector and then reduces its payments to interbank creditors, lowering the nominal amounts $\bar p_{ij}$ to actual payments $p_{ij} \leq \bar p_{ij}$.
As a result, adjacent nodes may also default, since the reduction in incoming payments decreases their asset sides.

\textbf{Clearing Principles in the Presence of External Shocks.} The framework proposed by Eisenberg and Noe~\cite{Eisenberg2001} introduces a simple yet powerful scheme for determining this ``fair'' payment matrix $P \doteq (p_{ij})$, based on three fundamental axioms: (i) limited liability, (ii) absolute priority of debt claims, and (iii) the pro-rata rule.
To formalize the axioms, we define the actual asset and liability sides of the banks, given a payment matrix $P$ and an external net position vector $c$:
\begin{equation}\label{eq:in-out-flows}
\phi^{\mathrm{in}} \doteq c^+ + P^\top \one, \quad
\phi^{\mathrm{out}} \doteq c^- + p, \quad \text{where } p \doteq P \one,
\end{equation}
representing, respectively, the total inflows and outflows at each node. The vector of residual values at the nodes, upon fulfilling external liabilities, is then given by
\begin{equation}\label{eq.vector-d}
d \doteq \phi^{\mathrm{in}} - c^- = c + P^\top \one.
\end{equation}

\noindent\textbf{(i) Limited liability.} The limited liability rule requires that the total payment of bank $i$ to the rest of the network should not exceed its net residual value. Mathematically, this means that bank $i$ fulfills the interbank debt claims only if $d_i>0$, and anyways its total payment $p_i$ cannot exceed $d_i$;
otherwise, bank $i$ is \emph{insolvent} and ceases all payments to other banks: $p_i=0$. These conditions can be written as
\begin{equation}\label{eq.limited-lia}
0\leq p_i\leq [d_i]^+\quad\forall i\in\calV.
\end{equation}
\textbf{(ii) Absolute priority of debt claims.} Each node $i$ either pays its obligations in full ($p_i=\bar p_i$) or pays all its value to the creditors ($p_i=[d_i]^+$). In view of the limited liability rule, this means that $p_i=\min(\bar p_i,[d_i]^+)$ for every bank $i\in\calV$.\vskip3mm
\noindent\textbf{(iii) Pro-rata (proportionality) rule.} Assuming that debts between the banks have equal seniority, it is natural to assume that the payments $p_{ij}$ from bank
 $i$ to its claimants $j\ne i$ have to be proportional\footnote{The proportional division principle is widely used in bankruptcy and tax legislation, is enforced in many financial contracts, and has been shown to be the only allocation (or division) rule satisfying several key axiomatic properties~\cite{Csoka2021}. Accordingly, we adopt this principle in line with the Eisenberg-Noe framework. However, while the pro-rata rule ensures local fairness in clearing payments between neighboring nodes, relaxing this rule can substantially reduce the system shortfall~\cite{Calafiore2024}.} to the nominal liabilities $\bar p_{ij}$. To write this formally,
it is convenient to introduce the stochastic\footnote{By definition, matrix $A$ is \emph{stochastic} if $a_{ij}\geq 0$ and $\sum_j a_{ij}=1$ for all $i$ or, equivalently, $A\one=\one$.} matrix of \emph{relative} liabilities
\begin{equation}\label{eq.A}
A=(a_{ij}),\quad
a_{ij}\doteq
\begin{cases}
\tfrac{\bar p_{ij}}{\bar p_i},& \bar p_i>0,\\
1, &\bar p_i=0 \mbox{ and } i=j,\\
0, &\text{otherwise.}
\end{cases}
\end{equation}
The {pro-rata (equal priority, proportionality) rule} can then be formulated as
\beq\label{eq:prorata}
p_{ij}=p_ia_{ij},\quad \forall i,j\in\calV,
\eeq
or, in matrix format, as $P = \diag(P\one)A$.

%\begin{remark}\rm
%The pro-rata rule typically guarantees uniqueness of the payment matrix and allows it to be found by solving an LP problem or through iterative algorithms, such as  the ``fictitious default'' algorithm~\cite{Eisenberg2001,Rogers2013}.
% It is important to note that this rule is more than a mere simplifying mathematical assumption. The proportional division principle is often implemented in bankruptcy and taxation legislature, enforced in many contracts and proves to be the only division rule satisfying several important properties~\cite{Csoka2021,Thomson2013,MorenoTernero2009}.
%However, while the pro-rata rule ensures local fairness of clearing between neighbor nodes, its lifting may significantly reduce the overall system loss (see Definition~\ref{def.loss})~\cite{Csoka2018,Calafiore2021a,Calafiore2024}.
%\erem
%\end{remark}

\textbf{Clearing Vectors and Incurred System Loss.} The three rules (i)-(iii) above can be written as a single nonlinear equation on the vector $p=P\one$: by virtue of the pro-rata rule~\eqref{eq:prorata},
$d=c+P^{\top}\one=c+A^{\top}p$, whence\footnote{Equation~\eqref{eq:clearing-2} is a special case of the so-called ``payment equilibrium'' equation from~\cite{Acemoglu2015}.}
\beq\label{eq:clearing-2}
p=\min(\bar p, [c + A\tran p]^+)=\left[\min(\bar p,c + A\tran p)\right]^+.
\eeq
%Since $\bar p\geq 0$, the latter equation can be alternatively written as

\begin{definition}\label{def.loss}
A vector $p$ satisfying ~\eqref{eq:clearing-2} is said to be
a \emph{clearing} vector for the financial network $(\calV,\calE,\bar P)$ with  vector of net incoming values $c$.
\end{definition}

Notice that if the actual payments $p_{ij}=a_{ij}p_i$ are determined by a clearing vector $p$, then the network's shortfall (the total reduction of interbank claims) is
found as
\begin{equation}\label{eq.loss-function}
L(p)\doteq\sum\nolimits_{i\in\calV}(\bar p_i-p_i)=\one^{\top}(\bar p-p).
\end{equation}
 \begin{definition}
We call $L(p)$ the overall {\em system loss} corresponding to clearing vector $p$.
\end{definition}

%In this study, our main focus lies in minimizing the worst-case system loss, which corresponds to the %most severe asset price variation of given magnitude. To cope with this problem, we employ specific %properties of clearing vectors outlined in the subsequent subsection.
%Clearing payments are associated to minimal system loss, as recalled in the following section.

In the \emph{generic} case, the clearing vector is known to be unique—this holds, for instance, when the network is strongly connected (a condition referred to as ``regularity'' in~\cite{Acemoglu2015}) and $\sum_{i \in \calV} c_i \ne 0$. When $c \geq 0$, uniqueness depends solely on the structural properties of the network~\cite{Calafiore2024,Csoka2024}. Necessary and sufficient conditions for uniqueness in the general signed case are provided in~\cite{Massai2022}. Even in cases where the clearing vector is not unique, a \emph{maximal} clearing vector $p^*(c)$ always exists -- one that minimizes the system loss function~\eqref{eq.loss-function} and can be computed efficiently. Its formal characterization is presented in the next section, which introduces the key method used to analyze worst-case shock scenarios in the subsequent analysis.

\subsection{Optimal Clearing Vectors and Loss Minimization}\label{sec:clearing-loss}

In this subsection, we present three propositions concerning the optimal (maximal) clearing vector, which is optimal in the sense that it minimizes the system loss. The proofs are deferred to Appendix.
%The first proposition is entailed by the well-known Knaster-Tarski fixed-point theorem~\cite{Elsinger2006,Glasserman2016}, however, we will give a simple direct proof in Appendix~\ref{sed:proof:prop:minmaxclearing}, which appears to be useful also in the next statements.
\begin{proposition}\label{prop:existence}
The set of clearing vectors is non-empty for any vector of net incoming values $c \in \mathbb{R}$. Moreover, there exist clearing vectors $p_*(c)$ and $p^*(c)$ that are, respectively, the componentwise \emph{minimal} and \emph{maximal} elements of this set, in the sense that any other clearing vector $p$ satisfies $p_*(c) \leq p \leq p^*(c)$.
\end{proposition}

Since the function $L(p)$ defined in~\eqref{eq.loss-function} is monotone decreasing\footnote{This means that $\tilde{L}(p') \geq \tilde{L}(p'')$ whenever $p',p''\in[0,\bar p]$, $p' \leq p''$ and $p'\ne p''$.} in $p$, the maximal clearing vector $p^*(c)$, obviously, provides the minimal overall system loss among all possible clearing vectors. Therefore, we assume henceforth that financial institutions adopt the \emph{maximal} clearing vector $p^*(c)$ and restrict our analysis to its properties.

The following proposition states that $p^*(c)$ can be computed by solving a linear programming problem, except in cases where severe drops in asset values make it impossible to avoid insolvency at certain nodes, that is, the equity of some financial institutions becomes negative even if they fully suspend their liabilities.

\begin{proposition}\label{prop:lp-feasible}
For given $c\in\Real{n}$,  nominal liability matrix $\bar P\in\Real{n,n}$, and stochastic pro-rata matrix $A$ given in \eqref{eq.A},
consider the linear program:
\bea
\eta^* =\min_{p\in\Real{n}} & \one\tran (\bar p - p)   \label{eq_clearingopt_LP}\\
 \mbox{s.t.:} & \bar p \geq p\geq 0 \nonumber \\
 & c + A\tran p  \geq p. \nonumber
 \eea
If this problem is feasible, then its optimal solution $p^*$ is unique
and coincides with the maximal clearing vector $p^*=p^*(c)$.
Moreover, under $p^*$, all liabilities towards the external sector are paid in full, i.e., vector $d$ in~\eqref{eq.vector-d} is nonnegative.
%These statements remain valid if the loss function $\mathbf{1}^{\top} (\bar p - p)$ in~\eqref{eq_clearingopt_LP} is replaced by any continuous decreasing function $\tilde{L}:[0,\bar p] \to \mathbb{R}$.

Contrary, if problem \eqref{eq_clearingopt_LP} is infeasible, then for any possible payment vector $0\leq p\leq \bar p$ (including the clearing vector defined via \eqref{eq:clearing-2}) there will exist nodes which are insolvent to the external sector, that is, $d_i=c_i+(A^{\top}p)_i<0$ and $p_i=0$ for some $i$.
 \end{proposition}

A proof of Proposition~\ref{prop:lp-feasible} is given in Appendix~\ref{sed:proof:prop:lp-feasible}.

%\begin{remark}\rm
%In the \emph{generic} scenario, the clearing vector is proven to be unique, thus $p_*(c)=p^*(c)$. This holds true, for example,
%when the network exhibits strong connectivity (referred to as ``regularity'' in the context of~\cite{Acemoglu2015,Ren2016})
%and $\sum_{i\in\calV}c_i\ne 0$~\cite{Acemoglu2015,Hurd2016}.
%In the case where $c\geq 0$, this uniqueness is solely contingent upon the structural characteristics of the graph~\cite{Calafiore2024}.
%The necessary and sufficient conditions for clearing vector uniqueness in the signed case
%are found in~\cite{Massai2022}, where also a complete characterization of clearing vectors is provided.

%We assume henceforth that the financial institutions are using the \emph{maximal} clearing vector $p^*(c)$, which both minimizes the system loss function and it is easy to compute. This vector can be computed via fixed-point iterations (see Appendix~\ref{sed:proof:prop:minmaxclearing}),
%or by solving an optimization
%problem (Proposition~\ref{prop:lp-feasible}). When $c\geq 0$, the maximal clearing vector can be determined through a modification
%of the ``fictitious default'' algorithm proposed in~\cite{Rogers2013}.
%\erem
%\end{remark}

%\vspace{.3cm}
Observe that the optimal value $\eta^*$ of \eqref{eq_clearingopt_LP} is  nonnegative;  it equals zero when no defaults occur and is positive otherwise, representing the total system-wide loss due to defaults. We adopt the convention of setting $\eta^* = \infty$ when problem~\eqref{eq_clearingopt_LP} is infeasible -- that is, when no payment vector $p$ satisfying $0 \leq p \leq \bar p$ ensures nonnegative equity for all participating banks. Moreover, the function $\eta^*s$ is \emph{convex} over the set $\calC_{feas}$ of vectors $c \in \mathbb{R}^n$ for which problem \eqref{eq_clearingopt_LP} is feasible, as established in the following.
\begin{proposition}\label{prop:eta-concave}
Let $\calC_{feas}$ be the set of all vectors $c\in\mathbb{R}^n$ for which the inequalities in~\eqref{eq_clearingopt_LP} have at least one feasible solution.
The maximal clearing vector $p^*$ is an element-wise non-decreasing and concave function on $\calC_{feas}$.
Symmetrically, the optimal value $\eta^*$ in~\eqref{eq_clearingopt_LP} is non-increasing and convex over $\calC_{feas}$.
\end{proposition}

A proof of Proposition~\ref{prop:eta-concave} is given in Appendix~\ref{sed:proof:prop:eta-concave}.
Notice that the set $\calC_{feas}$ is a convex closed polyhedral set, which is unbounded and contains all vectors $c\geq 0$.
However, if $c$ has negative components, it may happen that $\mathcal{C}_{\text{feas}}$ is empty, then some banks -- despite being fully relieved of interbank payment obligations (i.e., $p_i = 0$) -- still cannot meet their external liabilities without incurring negative equity, regardless of how an admissible payment vector $p \in [0, \bar{p}]$ is chosen.
% For the case $c\geq 0$, some statements of Proposition~\ref{prop:eta-concave} have been proved in~\cite[Lemma~5]{Eisenberg2001}, %which lemma, however, assumed additionally that the clearing vector is unique. We thus give the proof in Appendix that gets rid of this %constraint.

 %\vspace{.2cm}

\begin{figure}[h!tb]
  \centering
  \includegraphics[width=0.5\columnwidth]{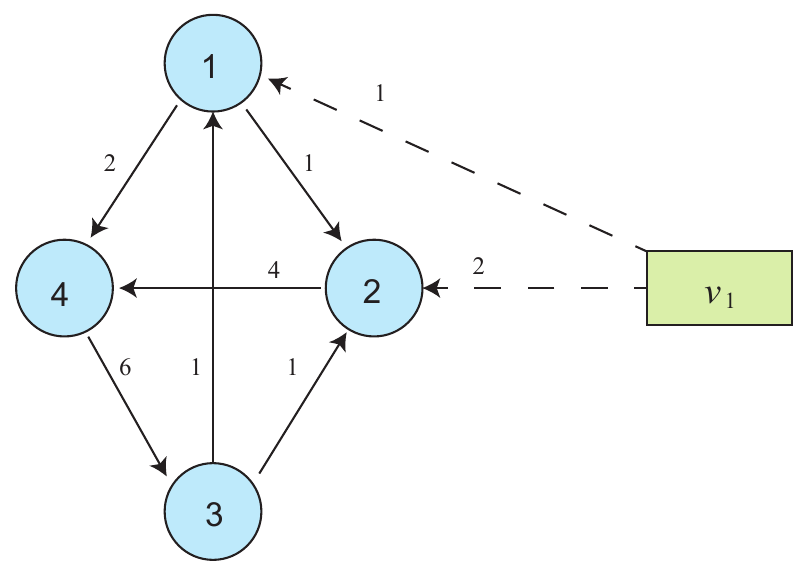}
  \caption{An illustrative example with $n=4$ nodes and one external asset.}\label{network_ex_toy}
\end{figure}

\paragraph{An illustrative example.}
\label{ex:toyex}
Figure~\ref{network_ex_toy} shows a small example with $n=4$ nodes,  mutual liabilities
$\bar p_{12}=1$, $\bar p_{14}=2$, $\bar p_{24}=4$, $\bar p_{31}=1$,
$\bar p_{32}=1$, $\bar p_{43}=6$, exposures in external assets $s_{11}=1$,   $s_{21}=2$ (only one external asset in this example),
and nominal value of the external asset $v=v_1=2.2$.
Assuming $c^e=0$ in (\ref{eq:extassets}), we have $c=Sv$, and solving problem~\eqref{eq_clearingopt_LP} we have that no default arises, whence $ \eta^*(2.2) = 0$, and
$p^* = \bar p = [3,\; 4,\; 2,\; 6]\tran$.
However, if the value of the external asset drops to $v = 1.9$, then $\eta^*(1.9) = 0.1667$
and the clearing vector results to be
$
p^* = [ 2.9,\;     4.0 ,\; 2.0,\;     5.93]\tran \leq \bar p\tran
$,
showing that nodes $1$ and $4$ default. If the value of the external asset drops  to $v = 1.5$
we have $\eta^*(1.5) = 0.8333$ and
$
p^* = [ 2.5,\;     4.0 ,\; 2.0,\;     5.6667]\tran \leq \bar p\tran$,
still with nodes $1$ and $4$ in default.
If $v$ drops further to $v=1$, then $\eta^*(1) = 2.333$ and $
p^* = [ 2.0,\;     3.6667 ,\; 2.0,\;     5.0]\tran \leq \bar p\tran$,
where now also node $2$ defaults.
The main contribution of this paper is to provide an efficient computational method for determining the threshold value on the amplitude of $v$ (in the more general case in which $v$ is a vector, rather than a scalar) that guarantees a no-default operation of the financial network, and for determining the loss curve of $\eta^*$ versus the perturbation level, once such threshold is surpassed.

In this example, there are no insolvencies with respect to the external sector, as all banks have zero net positions toward it.
If, on the other hand, one assumes that $c_{1}^e = -2$, meaning that the first node owes 2 units to the external sector, then this debt clearly cannot be repaid when $v\geq 1$. If the asset price $v$ falls below $1$, then bank~1 cannot fulfill its obligation to the external sector -- even if all interbank payments are suspended, since its inflow $\phi\ap{in}_1$ obviously does not exceed $s_{11}v+\bar p_{31}=1+v$.

\section{Main Results}\label{sec:main}

We henceforth assume that the net external position vector $c$ consists of a nominal component $\bar c$ and a variable component,
caused by the price fluctuations, that is,
\begin{equation}\label{eq.fluctuation1}
c = \bar c + S\delta,\quad \bar c\doteq \bar c^e + S\bar v,\quad v=\bar v +\delta
\end{equation}
where $\bar c^e$, $\bar v$ denote the nominal values of the net liquidity incoming from the external sector and of the asset prices, respectively, and $\delta\in\Real{m}$ is a vector of price fluctuations.
We recall that, in nominal conditions, the book values of the {asset} and {liability} sides of the balance sheets for all banks~\eqref{eq.in-out-nominal} can be written as
\begin{equation}\label{eq.in-out_nominal}
\bar\phi\ap{in} =  {\bar c}^+ + A\tran \bar p, \quad
 \bar\phi\ap{out}= {\bar c}^- + \bar p.
\end{equation}

\noindent
We consider the following assumption.
 \begin{assumption}[No defaults on interbank or external obligations under nominal conditions.]\rm
 \label{ass:nodefnom}
 $A$, $\bar p$, $S$, $\bar c^e$, $\bar v$ are such that
 \begin{equation}\label{eq.nominal-nodefault}
\bar c + A\tran \bar p > \bar p.
 \end{equation}
% \bar c= (\bar c^e + S\bar v )
That is, under nominal liabilities, nominal net external inflows and nominal asset prices, all banks can fulfill their interbank and external obligations. \est
\end{assumption}

\subsection{The Default Resilience Margin}\label{sec:primedefaults}

The first question we pose is about how much fluctuation in the assets' prices the system can withstand before a default is triggered in some bank.

   \begin{definition}[Default resilience margin]
   Under Assumption~\ref{ass:nodefnom},
the {\em  default resilience margin} of the banking system with respect to the norm $\|\cdot\|$ on the space of price fluctuations $\{\delta\}$ is defined as the maximum value $\epsilon^*$ of $\epsilon > 0$ such that
  \beq
\bar c + S\delta + A\tran \bar p \geq \bar p ,\quad \forall \delta: \|\delta\|\leq \epsilon.
 \label{eq:jarm}
 \eeq
 \end{definition}
 In other words, the resilience margin
 is  the maximum joint price perturbation amplitude that guarantees the system to remain default-free in the worst case.

 %We call {\em primary defaulters} those banks who default due to the immediate effect of the drop %in asset prices, that is

 \subsubsection{Computing the resilience margin}

The resilience margin is found by solving
 \begin{eqnarray}
 \epsilon^* = & \max \epsilon \\
 \mbox{s.t.: } &
\bar c + S\delta + A\tran\bar p \geq \bar p, &\; \forall \delta: \|\delta\|\leq \epsilon.\notag
 \end{eqnarray}
Due to symmetry in  $\|\delta\|\leq \epsilon$ we can replace $\delta = -\tilde \delta$
and rewrite the requirement as
\beq\label{eq.bar-r}
S\tilde \delta \leq \bar r \doteq    \bar c + (A\tran - I) \bar p, \quad \forall \tilde \delta : \|\tilde \delta\|\leq \epsilon,
\eeq
which can be rewritten as
\beq
\max_{\tilde \delta: \|\tilde \delta\|\leq \epsilon} \sigma_i\tran \tilde \delta \leq
%  [\bar c + (\bar P\tran - \bar P) \one]_i,
\bar r_i
\quad \forall i\in\calV
\label{eq:lhs_wc}
\eeq
where  $\sigma_i\tran$ denotes the $i$th row of $S$, and we notice that $\bar r_i$ is the nominal net worth of bank $i$, in the absence of price perturbations.
The actual solution now depends on the specific choice of the norm.
Introducing the \emph{dual} norm
\begin{equation}
\|\sigma\|_*\doteq \max_{\|\delta\|\leq 1}\sigma^{\top}\delta,
\end{equation}
the condition~\eqref{eq:lhs_wc} can be reformulated in the equivalent form
\beq\label{eq:lhs_wc+}
\epsilon\|\sigma_i\|_*\leq\bar r_i\quad\forall i\in\calV.%\min_{i\in\calV}\frac{\bar r_i}{}
\eeq
This leads to our first result, offering a simple formula for the default resilience margin.
\begin{thm}\label{thm:margin}
The default resilience margin with respect to the norm $\|\cdot\|$  is found as
\beq\label{eq.eps-star}
\epsilon^*\doteq\min_{i\in\calV}\frac{\bar r_i}{\|\sigma_i\|_*},
\eeq
where vector $\bar r$ is defined in~\eqref{eq.bar-r}. For any price perturbation $\delta$ such that $\|\delta\|\leq \epsilon^*$ the financial system remains default free. There exists, however, a \emph{worst-case} perturbation with $\|\delta\|=\varepsilon^*$, which brings the balance of some bank to zero.
\end{thm}

The proof of Theorem~\ref{thm:margin} proceeds straightforwardly. Define $\varepsilon^*$ as in~\eqref{eq.eps-star}. Then, inequalities~\eqref{eq:lhs_wc+} (and consequently, conditions~\eqref{eq.bar-r} and~\eqref{eq:lhs_wc}) become equivalent to the condition $\varepsilon\leq\varepsilon^*$. Let $\mathcal{I}^*\subseteq\mathcal{V}$ denote the set of indices on which the minimum in~\eqref{eq.eps-star} is attained. By selecting $i_0\in\mathcal{I}^*$, we can select a perturbation $\delta=-\tilde\delta_*$, where
\beq
\delta_*\doteq\varepsilon^*\arg\max\{\sigma^{\top}_{i_0}v:v\in\mathbb{R}^n,\|v\|=1\}.
\eeq
This choice effectively nullifies the balance of bank $i_0$, transforming the inequality~\eqref{eq:lhs_wc} with $i=i_0$ into an equality.\est

\subsubsection{Two relevant special cases}

Although the dual norm can easily be computed for any $\ell_p$ norm on $\mathbb{R}^n$, in this work we are primarily interested in two  cases that have a clear financial significance, namely the $\ell_\infty$ and the $\ell_1$ norm.
The $\ell_\infty$ constraint $\| \delta\|_\infty \leq \epsilon$ implies independent, entry-wise bounds on the variation of each asset price, that is $|\delta_i| \leq \epsilon$, for $i=1,\ldots,m$. The $\ell_1$ constraint $\| \delta\|_1 \leq \epsilon$ implies instead a constraint on the sum of absolute price perturbations, that is $\sum_i |\delta_i| \leq \epsilon$.

\vspace{.3cm}
\paragraph{$\ell_\infty$ case}

 For the $\ell_\infty$ norm $\|\cdot\|=\|\cdot\|_{\infty}$ the dual norm is $\|\cdot\|_*=\|\cdot\|_{1}$, and
\beq
\max_{\tilde \delta: \|\tilde \delta\|_\infty \leq \epsilon} \sigma_i\tran \tilde \delta =
\epsilon \|\sigma_i\|_1,
\eeq
 which maximum is attained for
 $\tilde \delta_j^* = \epsilon \cdot\mbox{sign}(S_{ij})$, $j=1,\ldots,m$.
The default resilience margin is thus
 \begin{equation}
 \epsilon^*_\infty =  \min_{i=1,\ldots,n} \frac{\bar r_i}{ \|\sigma_i\|_1}.
 \label{eq:res_marg_linfty}
 \end{equation}
and, correspondingly, a worst-case perturbation
 \begin{equation}
\delta^{(\infty)} = - \epsilon^*_\infty \cdot\mbox{sign}(\sigma_o),
 \label{eq:linf_wc}
 \end{equation}
with
  $o\in\calI^*$, where  $\calI^*$ denotes the set of indices $i\in\{1,\ldots,n\}$ for which the  minimum in \eqref{eq:res_marg_linfty} is attained.
The banks with $i\in\calI^*$ are called {\em primary defaulters} since their balance sheets are brought to zero (hence to the brink of default) by a critical  movement $\delta^{(\infty)}$ of the asset prices  of amplitude $\epsilon^*_\infty$.
Should the value of asset prices move beyond
 $\delta^{(\infty)}$ (positively, if $\delta^{(\infty)}_j >0$, or negatively if $\delta^{(\infty)}_{j} <0$)
 then all primary defaulters will actually default and, as a consequence of these defaults, they may trigger other {\em secondary defaults} in banks who see their balance sheet drop negative due to reduced income from primary defaulters, and so on.
Contrary, for any price perturbation $\delta$ such that $\|\delta\|_\infty\leq \epsilon^*_{\infty}$ the system is guaranteed to remain default free.

We observe that the worst-case $\ell_\infty$ price perturbations in \eqref{eq:linf_wc} represent a rather pessimistic situation in which all assets prices simultaneously drop (or rise, depending on the sign of the corresponding entry in $\sigma_o$) by the maximum margin
$\epsilon^*_{\infty}$. This aspect is mitigated by considering bounds on the joint variation of all asset prices, as captured by the $\ell_1$ norm of the price perturbations, as discussed next.

\vspace{.3cm}
\paragraph{$\ell_1$ case}
For the  $\ell_1$ norm case, the dual norm is the $\ell_{\infty}$ norm, and
\beq
\max_{\tilde \delta: \|\tilde \delta\|_1 \leq \epsilon} \sigma_i\tran \tilde \delta =
\epsilon \|\sigma_i\|_\infty.
\eeq
Letting ${\mathcal J}^*_i \doteq \arg\max_{j=1,\ldots,m} |S_{ij}|$, the above maximum is attained for
perturbations $\tilde \delta$ such that  all entries are zero except for those with indices
$j\in {\mathcal J}^*_i$, which take value $\pm\epsilon/|{\mathcal J}^*_i|$.
  The default resilience margin is given by
 \begin{equation}
\epsilon^*_1 =  \min_{i=1,\ldots,n} \frac{\bar r_i}{ \|\sigma_i\|_\infty}.
\label{eq:res_marg_l1}
 \end{equation}
Denoting again with $\calI^*$ the set of indices $i\in\{1,\ldots,n\}$ for which the  minimum in \eqref{eq:res_marg_l1} is attained,
by taking any $i\in\calI^*$
we obtain a corresponding worst-case price perturbation scenario in the $\ell_1$ case as a vector
$\delta^{(1)}$ in which all entries are zero, except for those in positions
$j\in{\mathcal J}^*_i$, which take value
\beq
\delta^{(1)}_j = -\frac{\epsilon^*_1}{|{\mathcal J}^*_i|} \mbox{sign}(S_{ij}),\quad \forall j\in {\mathcal J}^*_i.
\eeq
Observe that in common situations the optimal index sets $\calI^*$ and ${\mathcal J}^*_i$, $i\in\calI^*$,
will contain just one element, hence in such cases the optimal $\delta^{(1)}$ perturbation will consist of only one nonzero entry, and $\calI^*$, ${\mathcal J}^*_i$
identify the most critical bank and the most critical asset, respectively.

\vspace{.3cm}
In the next section we consider the situation in which the amplitude $\epsilon$ of the perturbation goes beyond the default resilience threshold $\epsilon^*$, and hence defaults may  appear. In such case, we are interested in determining the worst-case impact of the cascaded defaults on the system, i.e., in computing the  worst-case loss due to defaults.

\subsection{The worst-case loss curve}\label{sec:worst-case}
Consider problem \eqref{eq_clearingopt_LP} with $c = \bar c + S\delta$: its optimal value
$\eta^* = \eta^*(\bar c + S\delta)$ is a function of the price perturbation $\delta$, and we know that under Assumption~\ref{ass:nodefnom} we shall have $\eta^*(\bar c + S\delta)=0$ for all
$\delta$ such that $\|\delta\|\leq \epsilon^*$, where $\epsilon^*$ is the default resilience margin relative to the considered norm. As $\|\delta\|$ is allowed to go beyond  the $ \epsilon^*$ level, we shall have $\eta^*(\bar c + S\delta)\geq 0$ since defaults will be triggered.
We are here interested in computing the worst-case value of the system-wide financial loss $\eta^*(\bar c + S\delta)$ that can occur when $\|\delta\| \leq \epsilon$, for $\epsilon$ possibly larger than the resilience margin.
This is formalized as the following max-min problem
 \bea
\eta\ped{wc} =\max_{\|\delta\|\leq \epsilon }\eta^*(\bar c+S\delta) = \max_{\|\delta\|\leq \epsilon }\min_{p\in\Real{n}} & \one\tran (\bar p - p)   \label{eq_clearingopt_LP3}\\
 \mbox{s.t.:} & \bar p \geq p\geq 0 \nonumber \\
 & \bar c + S\delta + A\tran p  \geq p. \nonumber
 \eea
Its optimal value $\eta\ped{wc}$ would quantify the worst-case systemic impact of asset price variations in the range $\|\delta\|\leq \epsilon$.
The following key result holds.

\begin{thm}\label{prop:wc1}
The optimal value $\eta\ped{wc}$ of \eqref{eq_clearingopt_LP3} can be computed by solving the optimization problem
\bea
\eta\ped{wc} = \max_{\beta,\lam\geq 0}  & \; (\one -\beta)\tran \bar p
-  \bar c\tran  \lam +
 \epsilon \|S\tran \lam\|_*
 \label{eq_clearingopt_LPWC}\\
 \mbox{s.t.: } & \beta -\one + (I-A)\lam \geq 0\nonumber
 \eea
 where $\|\cdot\|_*$ is the dual of norm $\|\cdot\|$.
 \end{thm}

A proof of this theorem is given in Section~\ref{sed:proof:prop:wc1} in the Appendix.
Observe that, in general, this formulation can be challenging to solve numerically, as it involves the \emph{maximization}
of a convex function over a polyhedron, such problems are generally NP-hard.
%Such problems are  NP-hard in general; for example, this has been proven for
%$\ell_p$-norm maximization with $1 < p < \infty$~\cite{MangasarianShiau1986} and for many other convex and quasiconvex functions~\cite{bodlaender1990computational}.
In the next subsections, we show that the above problem actually admits
an efficient formulation  in the cases  where the price perturbation is bounded by either the $\ell_\infty$ or the $\ell_1$ norm.

 \subsubsection{Worst-case loss for $\ell_1$-norm price variations}
 \label{sec:wc_loss_l1}
 Consider first the case where the $\delta$ perturbation  is measured via the $\ell_1$ norm, $\|\delta\|_1$. In this case, the dual norm is the infinity norm, that is
 \[
  \|S\tran \lam\|_* = \|S\tran \lam\|_\infty = \max_{i=1,\ldots,m} |\zeta_i\tran \lam | =
  \max_{i=1,\ldots,m} |\zeta_i|\tran \lam ,
  \]
  where $\zeta_i\tran$, $i=1,\ldots,m$, are the rows of $S\tran$, and since  $\lam\geq 0$ it follows that $ |\zeta_i\tran \lam | = |\zeta_i|\tran \lam$.
Problem (\ref{eq_clearingopt_LPWC}) then becomes
\bea
\eta\ped{wc} = \max_{\beta,\lam\geq 0}  &\;  (\one -\beta)\tran \bar p - \bar c\tran \lam   +
 \epsilon \max_{i=1,\ldots,m} |\zeta_i|\tran \lam
 \label{eq_clearingopt_LP6}\\
 \mbox{s.t.: } &  \beta  - \one +(I -A)\lam \geq 0. \nonumber
 \eea
 The latter optimization problem can be rewritten as follows:
 \bea
\eta\ped{wc} = \max_{i=1,\ldots,m}  \max_{\beta,\lam\geq 0}  &  (\one -\beta)\tran \bar p -  \bar c\tran \lam   +
 \epsilon  |\zeta_i|\tran \lam
 \label{eq_clearingopt_LP7+}\\
 \mbox{s.t.: } & \beta   - \one +(I-A)\lam \geq 0. \nonumber
 \eea
 This means that we can compute the worst-case system loss efficiently and globally by solving $m$ linear programs, each of which amounts to solving the inner maximization in \eqref{eq_clearingopt_LP7+}
 with respect to $\beta,\lam$, for a fixed $\zeta_i$.
  Little further elaboration will give us also a worst-case price variation vector, and the clearing vector $p$ which is worst-case optimal.
 Let $\lam^*$ be an optimal $\lam$ vector for the above problem, then an $\ell_1$ worst-case price perturbation can be found as
$
 \delta^{(1)}_i
 $ such that
 \begin{equation}
 \label{eq:wc_l1}
  \delta^{(1)}_i =
  \left\{\ba{ll}
   - \epsilon\cdot \mbox{sign}( \zeta_i\tran \lam^* ) & \mbox{if } i= i^* \\
    0  & \mbox{otherwise},\quad
    i=1,\ldots,m,
  \ea\right.
      \end{equation}
      where $i^*$ is any index for which $|\zeta_i\tran \lam^*|$ is maximum.

      An interesting question arises about the uniqueness of the worst-case price perturbation
      $\delta^{(1)}$. This may be relevant,  since the
      worst-case  perturbation identifies the subset of assets whose price perturbation is the most critical and, correspondingly, the set of banks who will default due to such critical price fluctuation. Leveraging a known result of \cite{Mangasarian:979}, we provide in
      Proposition~\ref{prop:uniqueness} an easily computable criterion for checking whether the worst-case perturbation is unique.

 \subsubsection{Worst-case loss for $\ell_\infty$-norm price variations}
 \label{sec:wc_loss_linfty2}
When the $\delta$ perturbation  is measured via the $\ell_\infty$ norm, $\|\delta\|_\infty$, we have
 \beq
  \|S\tran \lam\|_* = \|S\tran \lam\|_1 = \sum_{i=1}^m |\zeta_i\tran \lam | =
   \sum_{i=1}^m |\zeta_i|\tran \lam  = \one\tran |S|\tran \lam
  \eeq
Problem (\ref{eq_clearingopt_LPWC}) then becomes
\bea
\eta\ped{wc} = \max_{\beta,\lam\geq 0}  & \; (\one -\beta)\tran \bar p -  \bar c\tran \lam   +
 \epsilon  \one\tran |S|\tran \lam
 \label{eq_clearingopt_LP7++}\\
 \mbox{s.t.: } & \beta  - \one +(I-A)\lam \geq 0. \nonumber
 \eea
That is, in this case we can find exactly and efficiently the worst-case loss by solving a single LP.
A worst-case price perturbation can then be found as
$
 \delta^{(\infty)}
 $ such that
 \beq
  \delta^{(\infty)}_i =
   - \epsilon\cdot\mbox{sign}( \zeta_i\tran \lam  ) ,\quad i=1,\ldots,m.
\label{eq:wc_linf}
\eeq

An easily computable criterion for checking whether   such worst-case perturbation is unique is provided by the following proposition.
\begin{proposition}[Uniqueness of the worst-case perturbation scenario]\label{prop:uniqueness}
For  the $\ell_1$-norm case:
the worst-case perturbation scenario in eq.\ \eqref{eq:wc_l1}
is unique if the conditions hold:
\begin{itemize}
\item the max in eq.\ \eqref{eq_clearingopt_LP7+} is attained at a single index $i$;
\item the corresponding maximization problem in $(\beta,\lam)$ has a unique solution
$(\beta^*,\lam^*)$   (a fact that can be checked by applying Proposition \ref{prop_unique});
\item $\max_i |\zeta_i\tran \lam^*|$ is attained at a single index $i$.
\end{itemize}
Similarily, for the $\ell_\infty$-norm case:
the worst-case perturbation scenario in eq.\ \eqref{eq:wc_linf}
is unique if the following two conditions hold:
\begin{itemize}
\item the  maximization problem \eqref{eq_clearingopt_LP7++} has a unique solution $(\beta^*,\lam^*)$
 (a fact that can be checked by applying Proposition \ref{prop_unique});
\item $ \zeta_i\tran \lam^*$ is nonzero for all $i$.
\end{itemize}
\end{proposition}
\color{black}

\subsection{The Insolvency Resilience Margin}\label{sec:upperlimit}

From (\ref{eq_clearingopt_LPWC}), we observe that the optimal worst-case loss value $\eta_{\mathrm{wc}}$ is a function of the perturbation level~$\epsilon$. As established in Section~\ref{sec:primedefaults}, we have $\eta_{\mathrm{wc}}(\epsilon) = 0$ for all $\epsilon \leq \epsilon^*$, where $\epsilon^*$ denotes the default resilience threshold under the chosen norm. Beyond this threshold, $\eta_{\mathrm{wc}}(\epsilon)$ begins to increase as $\epsilon$ grows. However, once the perturbation set ${\delta : |\delta| \leq \epsilon}$ becomes sufficiently large, it may include some vector $\delta$ for which Problem~\eqref{eq_clearingopt_LP3} becomes infeasible, i.e., $\bar c + S\delta\not\in \mathcal{C}\ped{\mathrm{feas}}$. In this section, we compute the largest perturbation level $\epsilon$ such that Problem~\eqref{eq_clearingopt_LP3} remains feasible for all $\delta$ with $|\delta| \leq \epsilon$, called the \emph{insolvency resilience margin}.

\begin{definition}[Insolvency resilience margin]
   The {\em  insolvency resilience margin} $\epsilon\ped{ub}$
   of the banking system with respect to the norm $\|\cdot\|$
   is the largest value of $\epsilon \geq 0$ such that
   the set $\{p:
    \bar p \geq p\geq 0,\; \bar c+ S\delta + A\tran p  \geq p\}$
    remains nonempty for all
    $\delta: \|\delta\|\leq \epsilon$.
 \end{definition}

The loss curve of $\eta\ped{wc}$ as a function of $\epsilon$ is thus properly defined in the range
   $[0, \epsilon\ped{ub}]$. Notice that $\eta_{wc}$, obviously, is non-decreasing and, furthermore, it can be easily derived from Proposition~\ref{prop:eta-concave}
   %(the convexity of $\eta^*$)
   that $\eta_{wc}(\epsilon)$ is convex on $[0,\epsilon\ped{ub}]$.
   The insolvency resilience margin $\epsilon\ped{ub}$ can be computed according to the following theorem.
   \begin{thm}\label{prop:eps_ub}
   $\epsilon\ped{ub}$ is given by the optimal value of  the following LP:
   \beas
  \epsilon\ped{ub}= \max_{p,\epsilon} & \epsilon \\
   \mbox{s.t.:} & 0\leq p \leq \bar p \\
   & \bar c -\epsilon s + A\tran p \geq p,
   \eeas
   where $s$ is a vector such that $s_i \doteq \|\sigma_i\tran\|_*$, $i=1,\ldots,n$.
\end{thm}

Theorem~\ref{prop:eps_ub} is proven in Section~\ref{sec:proof:eps_ub} of the Appendix.

Technically, for $\epsilon > \epsilon_{\mathrm{ub}}$, there exist perturbations $\delta$ with $|\delta| \leq \epsilon$ such that the constraint $(\bar c + S\delta) + A^\top p \geq p$ in problem~\eqref{eq_clearingopt_LP3} cannot be satisfied by any $p \geq 0$. In such cases, the clearing vector can no longer be characterized as the optimal solution of~\eqref{eq_clearingopt_LP}. As discussed in Section~\ref{sec:clearing-loss}, this situation corresponds to the inability to clear interbank liabilities without revealing the insolvency of certain banks, whose obligations to the external sector exceed the value of their assets.
	
	Dealing with the case $\epsilon > \epsilon_{\mathrm{ub}}$ would require a formulation incorporating constraints of the form
\begin{equation}\label{eq:constr-insolvency}
	p\leq \max\left(\bar c + S\delta + A^\top p,0\right),
\end{equation}
which would introduce nonlinearity and destroy the convexity of the problem.
In this case, the extremal problem for determining the maximal clearing vector becomes a mixed-integer linear program (MILP), rather than a linear program; see the relevant formulation in~\cite{AraratMeimanjan2023}. Since our method for computing the worst-case loss relies on linear programming duality, it does not extend to MILPs. Consequently, evaluating the worst-case loss for $\epsilon > \epsilon_{\mathrm{ub}}$ lies beyond the scope of this manuscript.

\color{black}

%\iffalse
%      \subsection{The loss curve}
%We may set level $\epsilon$ as a percentage of the amplitude of the nominal prices $\bar v$, that is
%   \[
%  \epsilon =  \tilde \epsilon \|\bar v\|,
%\]
%where $ \tilde \epsilon\in(0,1)$ is the relative price perturbation level.
% So we may consider the function $\eta\ped{wc}(\tilde \epsilon)$ and plot the curve of
% $\eta\ped{wc}$ versus $\tilde \epsilon$.

%  We expect to have  $\eta\ped{wc}=0$ for all
% values of $\epsilon =  \tilde \epsilon \|\bar v\| $ that are less than or equal to the resilience margin (in the selected norm). Then, $\eta\ped{wc}$ will grow progressively as the perturbation level goes beyond the margin. This curve, obtained numerically by solving  repeatedly problem  (\ref{eq_clearingopt_LPWC}) for many grid values of $\tilde \epsilon$, is likely to be non-smooth, with kinks corresponding to perturbation levels that trigger different sets of defaulting banks (this to be verified by numerical experiments).
%\fi

\section{Numerical examples}\label{sec:numeric}

To illustrate the proposed approach numerically, we present two examples. The first involves a small network with 8 nodes, allowing for transparent analysis and interpretation. The second considers a larger network with a core–periphery structure, a topology commonly used to represent financial networks~\cite{hu2024robust,Craig2014,JieMa2025}.

\begin{figure}[h!tb]
  \centering
  \includegraphics[width=0.45\columnwidth]{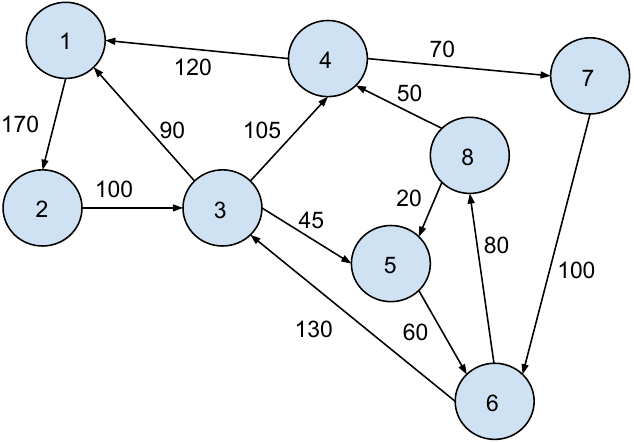}
  \caption{A schematic network with 8 nodes.}\label{network_ex}
\end{figure}

\textbf{Example 1.} First, we considered a schematic network with $n=8$ nodes and $m=4$ external assets, as shown in Figure~\ref{network_ex}, where the numbers on the arrows represent the mutual liabilities among the banks.
We assumed the vector of external inflows
$c^+$, external outflows $c^-$,
and the matrix of asset shares $S$ to be as follows:
\[
%\bar c=\left[\begin{smallmatrix}
%	-22\\
%	-62\\
%	15\\
%	140\\
%	25\\
%	130\\
%	40\\
%	-5
%	\end{smallmatrix}\right],
c^+=\begin{bmatrix}158\\
38\\
15\\
285\\
25\\
180\\
60\\
55\\
\end{bmatrix},\quad
c^-=\begin{bmatrix}180\\
100\\
0\\
145\\
0\\
50\\
20\\
60\\
\end{bmatrix},
\quad
S = \left[
%\begin{smallmatrix}
\begin{matrix}
96 & 29 & 99 & 0\\
53 & 13 &75 & 85\\
0 & 28 & 0 & 57\\
32 & 79 & 0 & 0\\
0 & 0 & 21 & 0\\
0 & 45 & 27 & 71\\
0 & 0 & 49 & 0\\
75 & 79 & 5 & 41\\
%\end{smallmatrix}
\end{matrix}
\right].
\]

For this network, we can  compute the resilience margins $\epsilon^*_\infty$ and
$\epsilon^*_1$, as given in  \eqref{eq:res_marg_linfty}
and in \eqref{eq:res_marg_l1} for the  $\ell_\infty$ case and for the $\ell_1$ case, respectively;
we also computed the perturbation upper limit $\epsilon\ped{ub}$ as shown in
Theorem~\ref{prop:eps_ub}, obtaining
\[
\epsilon^*_\infty= 0.0249,\quad
\epsilon^*_1 = 0.0630, \quad
\epsilon\ped{ub,\infty} = 0.1878,\quad
\epsilon\ped{ub,1} = 0.4328.
\]

We next selected 20 evenly spaced values of $\epsilon$ inside the interval $[\epsilon^*,\epsilon\ped{ub}]$ and for each of them we computed the worst-case loss as shown in Section~\ref{sec:wc_loss_l1} for the $\ell_1$ case and in Section~\ref{sec:wc_loss_linfty2} for the $\ell_\infty$ case.
For each value of $\epsilon$, we also modeled a financial shock by generating a random vector of price fluctuation $\delta<0$ such as $\|\delta\| =\epsilon$ and we computed the corresponding loss $\eta^*$ by solving Problem \ref{eq_clearingopt_LP}.

\begin{figure}[htb]
  \centering
  \begin{subfigure}[t]{0.5\textwidth}
      \includegraphics[width=\textwidth]{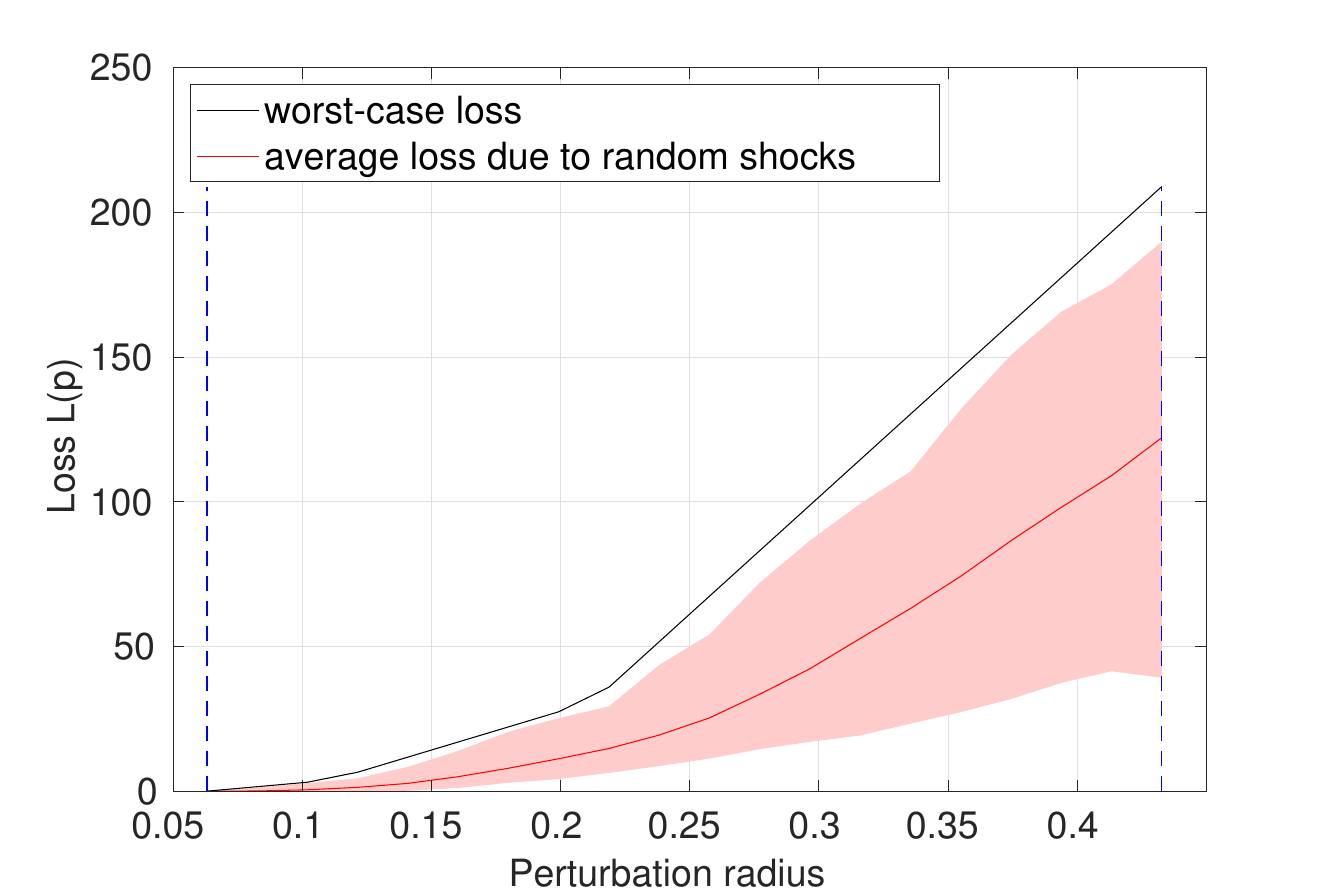}
  \caption{$\ell_1$-case}\label{fig:res_l1}
  \end{subfigure}\hfill
  \begin{subfigure}[t]{0.5\textwidth}
  %\centering
    \includegraphics[width=\textwidth]{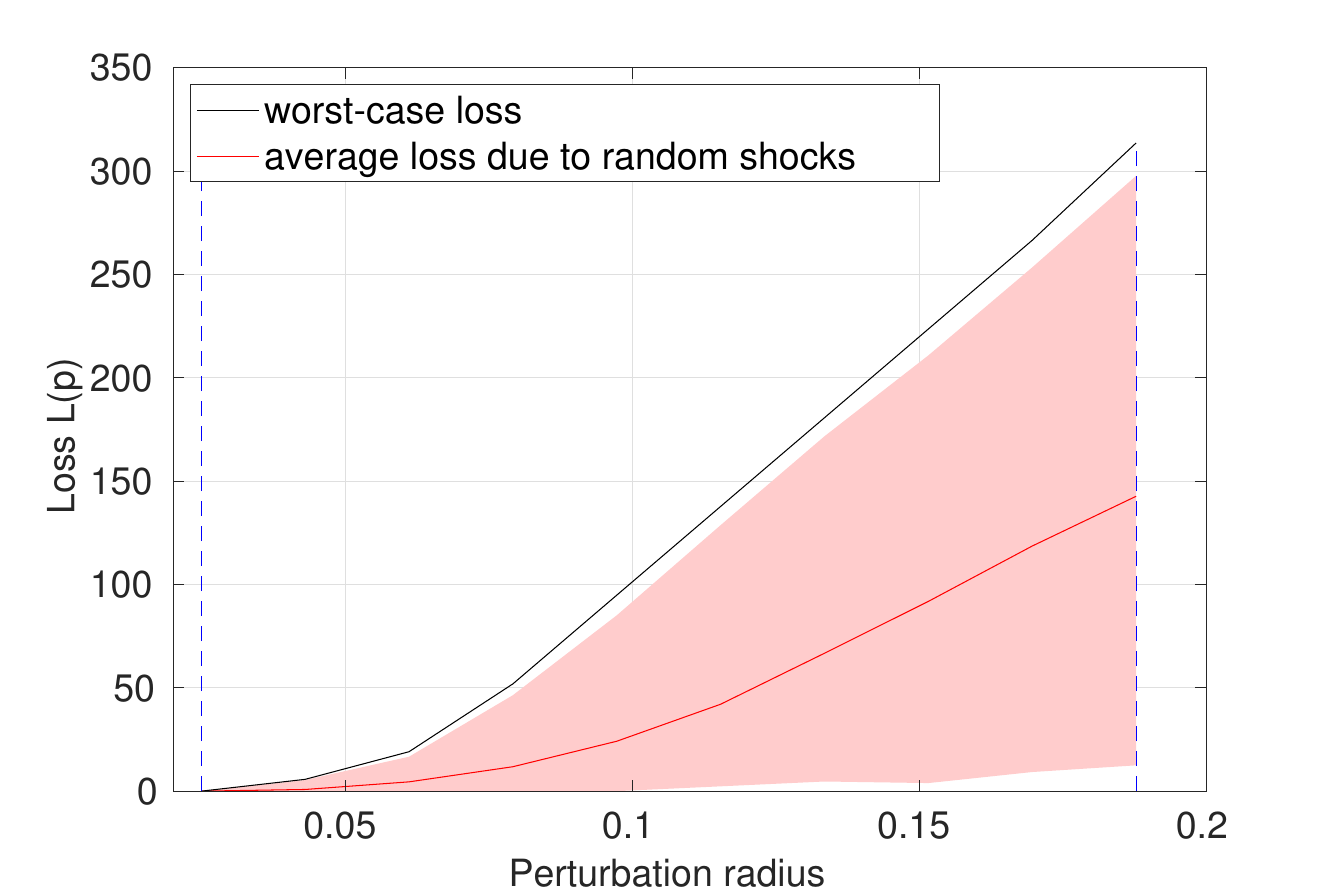}
  \caption{$\ell_{\infty}$-case}\label{fig:res_linf}
  \end{subfigure}
  \caption{Worst-case losses vs. losses under random shocks, $\ell_1$ and $\ell_\infty$ norms. The dashed lines mark the default resilience margin $\epsilon^*$ and the insolvency resilience margin $\epsilon\ped{ub}$.}
\end{figure}

In order to compare the worst possible losses to those produced by random shocks,
we performed 1000 runs for each value of $\epsilon$ and in each run we generated a different random scenario $\delta$. Figure \ref{fig:res_l1} and Figure \ref{fig:res_linf} show the obtained results for the $\ell_1$ case and the $\ell_\infty$ case, respectively. The red curve represents the average loss due to random shocks, and the red band represents, for each value of $\epsilon$, the interval between the minimum and maximum losses obtained in the random numerical simulation.

\begin{figure}[htb]
  \centering
  \includegraphics[width=0.5\columnwidth]{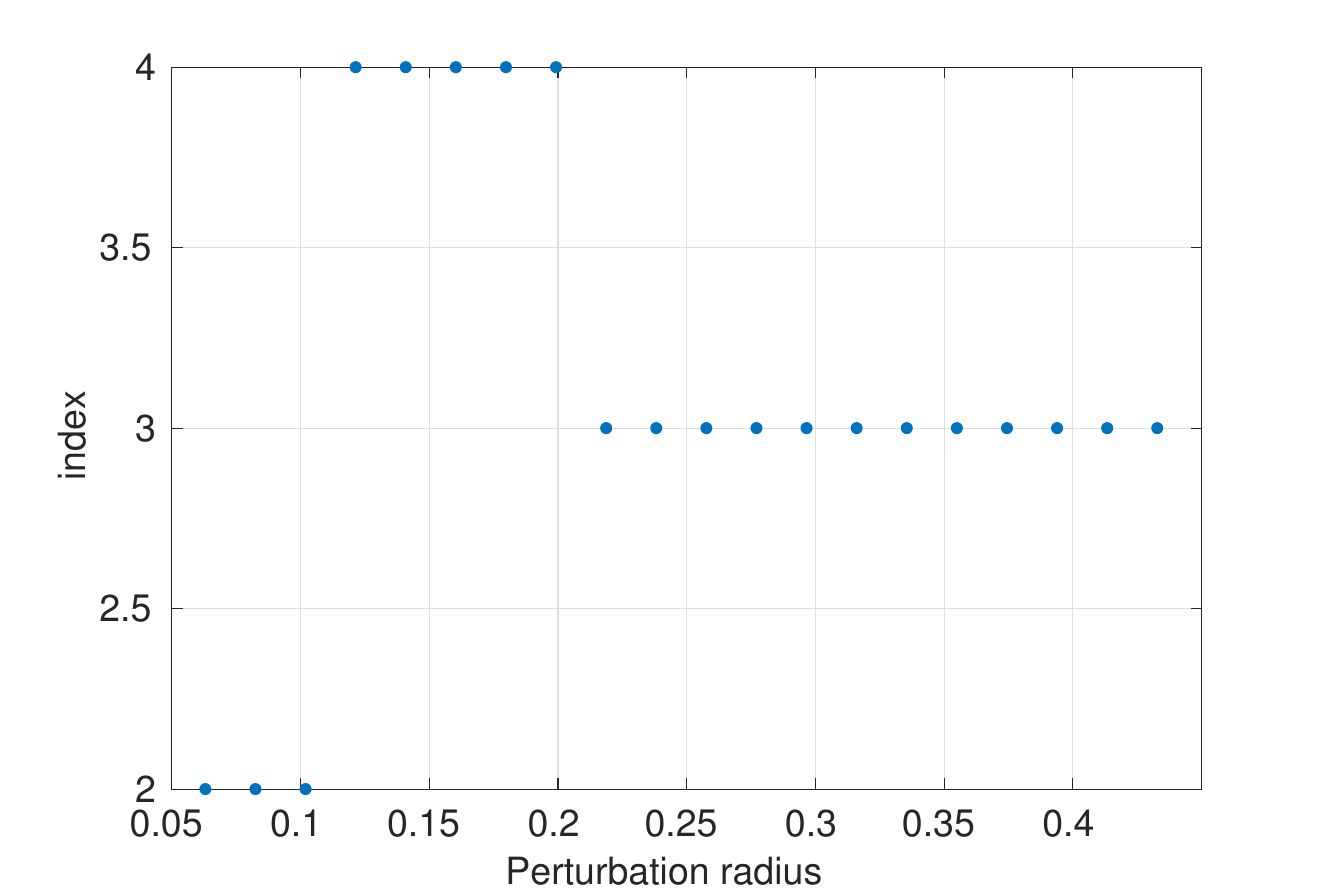}
  \caption{Index $i^*$ of the worst-case case perturbation $\delta^*$ in the $\ell_1$ case. Using Proposition \ref{prop:uniqueness}, we have checked that $\delta^*$ is unique for all $\epsilon>\epsilon^*$.
%  \cred{[GC: a different style of plot is needed here, eg. a stairs plot, to avoid slopes in the lines.]}
}\label{fig:imax}
\end{figure}

It is interesting to observe that the worst-case loss curve presents some ``kinks'', i.e., there are $\epsilon$ levels at which the slope of the curve changes abruptly. Intuitively, this feature is due to the change in the group of defaulting banks  as the level of perturbation increases.
%Mathematical analysis of LP problems, where the cost function depends on a parameter
%as in~\eqref{eq_clearingopt_LP6} and~\eqref{eq_clearingopt_LP7++}, is available %in~\cite[Chapter~V]{Gal1995}.
In \eqref{eq:wc_l1} we have shown that the worst-case price perturbation $\delta^*$ in the $\ell_1$ case is a vector where the only non-zero component corresponds to the asset $i^*$, defined as the one that maximizes $|\zeta_i\tran \lam|$. The asset $i^*$ is thus the one whose variation causes the worst-case impact.
Figure \ref{fig:imax} shows the value of $i^*$ as function of the perturbation radius $\epsilon$. By comparing Figure \ref{fig:res_l1} and \ref{fig:imax}, we can observe that the kinks observed in Figure \ref{fig:res_l1} are in correspondence of the changes of values of $i^*$.

\begin{figure}[htb]
  \centering
  \begin{subfigure}{0.5\textwidth}
    \includegraphics[width=\textwidth]{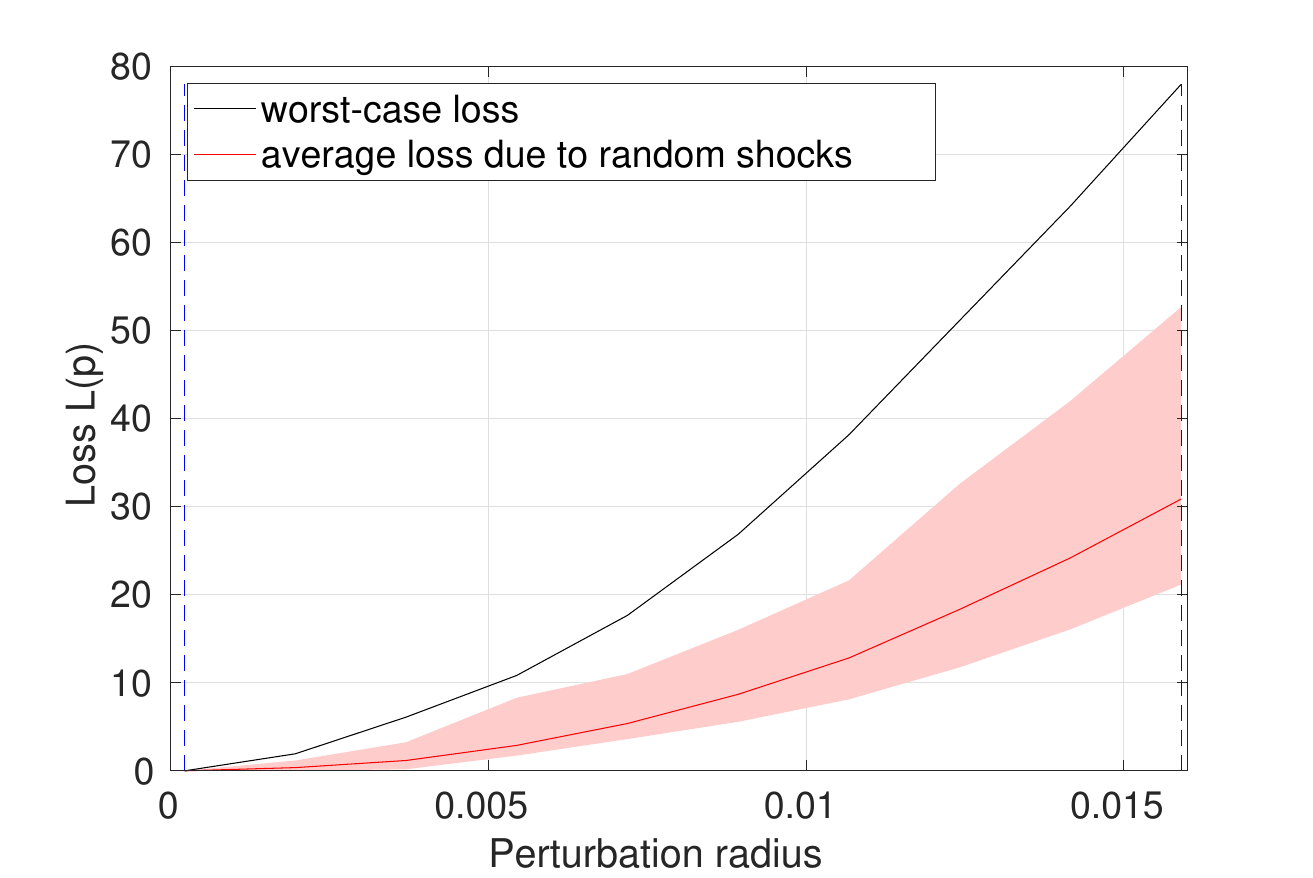}
  \caption{$\ell_1$-case}\label{fig:res_l1_coreper}
  \end{subfigure}\hfill
  \begin{subfigure}{0.5\textwidth}
  \centering
  \includegraphics[width=\textwidth]{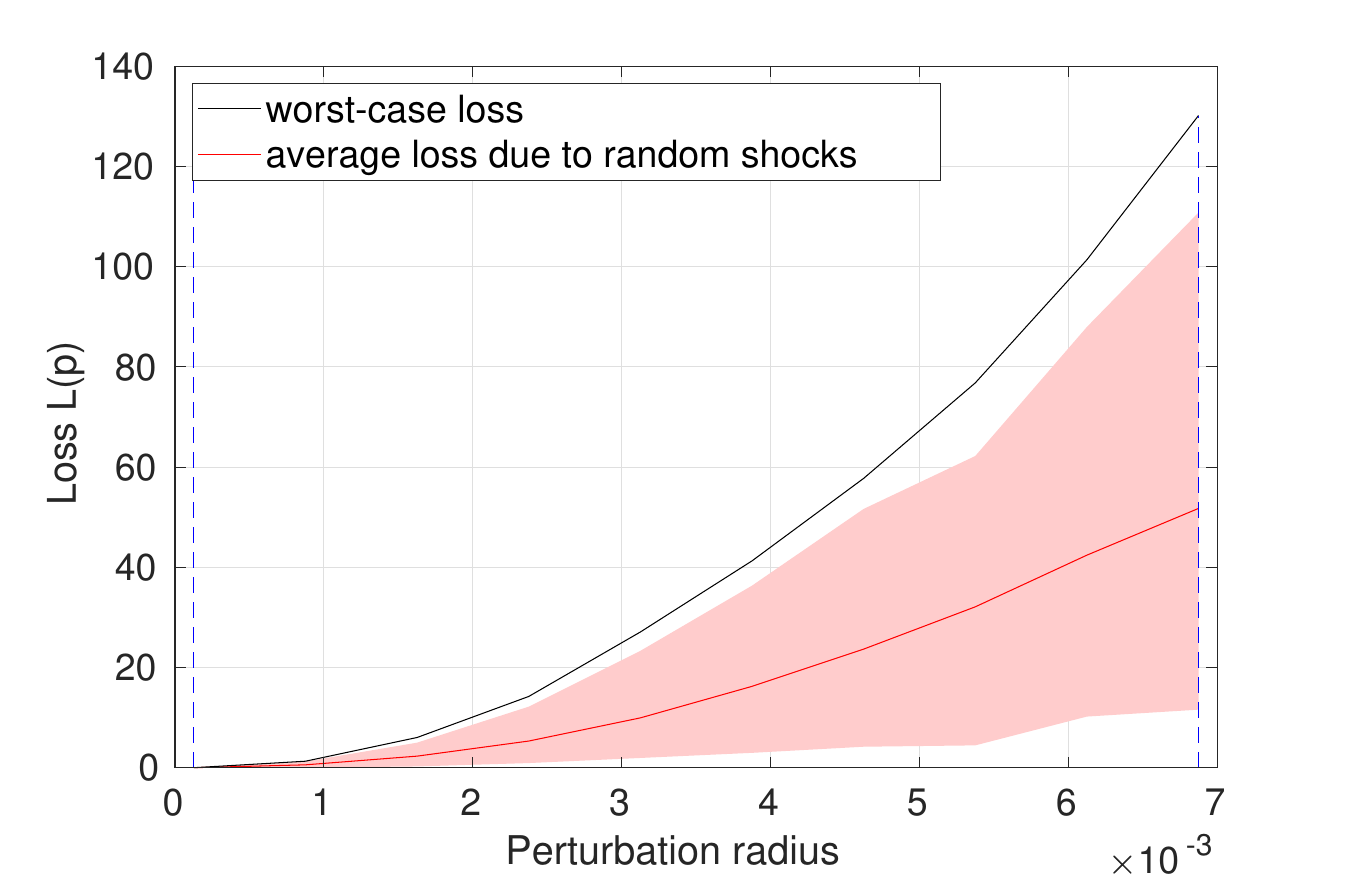}
  \caption{$\ell_{\infty}$-case}\label{fig:res_linf_coreper}
  \end{subfigure}
  \caption{Core-periphery random graph: Worst-case losses vs. losses under random shocks for the $\ell_1$ and $\ell_\infty$ norms. The dashed lines mark the default resilience margin $\epsilon^*$ and the insolvency resilience margin $\epsilon\ped{ub}$.}
\end{figure}

\begin{figure}[htb]
  \centering
  \includegraphics[width=0.5\columnwidth]{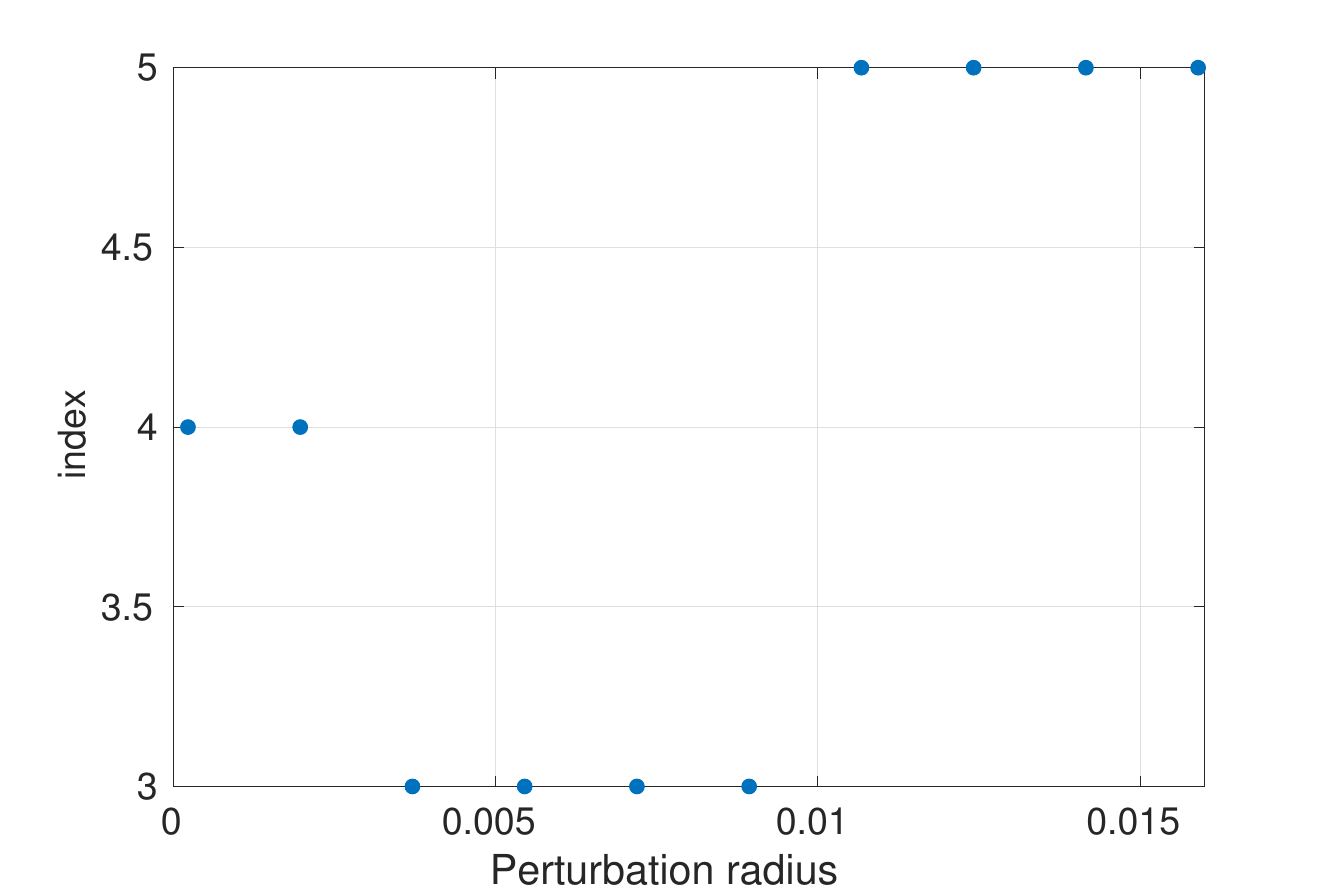}
  \caption{Core-periphery random graph: index $i^*$ of the worst-case case perturbation $\delta^*$ in the $\ell_1$ case. Using Proposition \ref{prop:uniqueness}, we have checked that $\delta^*$ is unique for all $\epsilon>\epsilon^*$.
%  \cred{[GC: a different style of plot is needed here, eg. a stairs plot, to avoid slopes in the lines.]}
}\label{fig:imax_coreper}
\end{figure}

\textbf{Example 2.} We now consider a core–periphery random graph with $n = 353$ nodes and $m = 5$ external assets, designed to mimic the structure of a financial network presented in~\cite{hu2024robust} (the details are given in Appendix \ref{sec:appendix:rndgraphconstr}). Similarly to the previous example, we select 10 evenly spaced values of $\epsilon$ inside the interval $[\epsilon^*,\epsilon\ped{ub}]$ and for each of them we compute the worst-case loss for the $\ell_1$ case and the $\ell_{\infty}$ case. Then, for each value of $\epsilon$, we can model a financial shocks by generating a random vector of price fluctuation $\delta<0$ such as $\|\delta\| =\epsilon$ and compute the corresponding loss $\eta^*$ by solving~\eqref{eq_clearingopt_LP}. We perform 1000 runs for each value of $\epsilon$, generating a different random scenario $\delta$ for each run. Figure \ref{fig:res_l1_coreper} and Figure \ref{fig:res_linf_coreper} show the obtained results for the $\ell_1$ case and the $\ell_\infty$ case, respectively. Figure \ref{fig:imax_coreper} shows the value of $i^*$ of the worst-case case perturbation $\delta^*$ in the $\ell_1$ case as function of the perturbation radius $\epsilon$.\color{black}

%\clearpage
\section{Discussion and Conclusions}\label{sec:conclusion}

We observe that two distinct but interrelated network effects contribute to financial contagion. First, the network of mutual liabilities creates direct contagion channels: when a bank defaults, it may fail to fully repay its obligations to a creditor bank, potentially triggering a chain reaction of defaults as each affected bank passes on reduced payments to its counterparties. Second, a separate network of common asset exposures links banks indirectly: when the market value of an external asset drops, all banks holding long positions in that asset (i.e., those with $s_{ij} > 0$) simultaneously suffer losses, proportional to their exposure levels. As a result, even a shock to a single asset can reduce multiple entries in the net external position vector $c$. A similar mechanism applies in the case of price increases for banks with short exposures.

Based on an extension of the classical financial network model of~\cite{Eisenberg2001, Elsinger2006}—incorporating common exposures to a set of external assets—this paper focused on two central questions: (i) what is the largest asset price perturbation the network can withstand without triggering defaults, and
(ii) what is the worst-case systemic loss incurred when perturbations exceed that threshold?
To this end, and allowing for sign-unrestricted external cash flows $c$, we first characterized the maximal clearing vector through a linear program (Proposition~\ref{prop:lp-feasible}). In Section~\ref{sec:primedefaults}, we introduced the concept of the \emph{default resilience margin} -- the maximum amplitude of asset price fluctuations the system can absorb without triggering any defaults. We showed how to compute this threshold under both the $\ell_\infty$ norm (maximum individual fluctuation) and the $\ell_1$ norm (total fluctuation magnitude), providing a rigorous analytical measure of the network’s \emph{robustness} to external shocks.
Section~\ref{sec:worst-case} explored the scenario where the perturbation exceeds the default resilience margin, leading to defaults. In this setting, we evaluated the worst-case systemic loss and demonstrated that it can be computed efficiently -- either by solving a single linear program in the $\ell_\infty$ case or a sequence of $m$ linear programs in the $\ell_1$ case -- along with identifying the corresponding worst-case shock scenario.
To visualize the system’s vulnerability against shocks,
we proposed the construction of the \emph{worst-case loss curve} $\eta_{\mathrm{wc}}(\epsilon)$, which maps the magnitude of asset price perturbation $\epsilon$ to the corresponding worst-case systemic loss.

Our analysis applies to perturbations up to an upper limit $\epsilon_{\mathrm{ub}}$, which is defined and characterized in Section~\ref{sec:upperlimit}. As discussed in that section, shocks with magnitude $\epsilon > \epsilon_{\mathrm{ub}}$ may result in the insolvency of some banks, preventing the determination of the maximal clearing vector through a linear program. Moreover, from a financial modeling perspective, scenarios involving insolvency with respect to the external sector should incorporate procedural aspects of firm liquidation and bankruptcy costs~\cite{Weber2017}. The computation of the worst-case losses in financial networks with defaults and bankruptcy costs~\cite{Banerjee2022,AraratMeimanjan2023} requires methodological instruments that are beyond the scope of the present paper and remains an important topic for future research.

 Another challenging problem is computing the worst-case loss curve under more general clearing mechanisms—such as those designed for financial markets with multiple clearing centers or central counterparties (CCPs)~\cite{Benos2024,Veraart2025,Tang2025}, as well as net liability clearing mechanisms~\cite{Ma2024}, which consolidate all bilateral obligations between pairs of financial institutions into net positions.

\iffalse
While the vector of price perturbation $\delta$ remains uncertain, we still assume that its maximum norm $|\delta|$ can be estimated. In the case of an illiquid asset, the reduction in its price hinges on the count of banks that participate in the ``fire sale'' of the asset, as well as their respective holdings in it. This complexity makes it challenging to provide an a priori estimate of the extent of the price decline.
\fi
Also, our model assumes the external assets to be liquid. Financial networks where the institutions have common exposures to \emph{illiquid} assets are described by more advanced models than the classical Eisenberg-Noe model (see, e.g.,~\cite{Cifuentes2005,Feinstein2017}) that revise the definition of a clearing vector and take liquidation costs into account. Even more complex interconnections between assets and liabilities are possible. Some recent models~\cite{Altermatt2024} examine self-fulfilling bank runs, where the value of bank assets hinges on households' ability to redeem their deposits. In these settings, liquidity misallocations -- where cash is withdrawn by households not intending to consume - undermine firm revenues and, consequently, bank asset returns, reinforcing the panic. A bank run occurs when many depositors simultaneously try to withdraw funds out of fear of failure. Since banks hold only a fraction of deposits as liquid reserves, such withdrawals can trigger or accelerate insolvency. Robustness of such models against external (market-driven or liquidity-driven) assets' price fluctuations remains, to the best of our knowledge, an open problem. Note that these contagion effects reflect higher-order interactions within the network~\cite{Boccaletti2023}, involving multiple nodes simultaneously and not reducible to simple pairwise relationships.

\section*{Data Availability declaration}
Data used in the numerical examples has been created randomly. It can be  reproduced by following the examples' description in the paper.

\section*{Funding and competing interest declaration}

The work has been supported by the project 2022K8EZBW ``Higher-order interactions in social
dynamics with application to monetary networks'', funded by European Union -- Next Generation EU within the
PRIN 2022 program (D.D. 104 - 02/02/2022 Ministero dell'Universit{\`a} e della Ricerca). This manuscript reflects
only the authors' views and opinions, and the Ministry cannot be considered responsible for them.
The authors  have no conflicts of interest to disclose.

\section*{Ethics, Consent to Participate, and Consent to Publish declarations}
Not applicable.

 \section*{Author Contribution declaration} The authors contributed equally to this work.

\clearpage
\appendix

\centerline{
{\Large  \bf Appendix (Supplementary Material)}}
\section{Proof of Proposition~\ref{prop:existence}}\label{sec:proof:prop:minmaxclearing}

Usually, Proposition~\ref{prop:existence} is derived from the Knaster-Tarski fixed point theorem~\cite{Eisenberg2001}. We give a direct proof,
which also gives a simple iterative procedure to compute the maximal and minimal clearing vectors.

We begin with some preliminary constructions. A clearing vector may be considered as a fixed point $p=T_c(p)$ of the monotone non-decreasing mapping $T_c:[0,\bar p]\to [0,\bar p]$, defined by the right-hand side of~\eqref{eq:clearing-2}, that is,
\[
T_c(p)\doteq\min\left(\bar p,[c+A^{\top}p]^+\right)=\left[\min(\bar p,c + A\tran p)\right]^+.
\]
Since $A$ is a nonnegative matrix, it is easily shown that $T_c:[0,\bar p]\to [0,\bar p]$ is \emph{non-decreasing} with respect to the ordering $\leq$.
Consider now the sequence of vectors:
\begin{gather}
  p^0(c)=\bar p,\;\; p^k(c)=T_c(p^{k-1}(c))\in [0,\bar p],\quad \forall k\geq 1\label{eq.iter-max-clearing}.
\end{gather}
Since $\bar p=p^0\geq p^1$, one has $p^1=T_c(p^0)\geq p^2=T_c(p^1)$ and, using the induction on $k$, one proves that $p^k\geq p^{k+1}$. Also, for each fixed $c$, the map $T_c$ is continuous. Passing to the limit as $k\to\infty$, one proves that
$p^*(c)\doteq\lim_{k\to\infty} p^k$ is a fixed point of $T_c$, that is, a clearing vector.
We have proved that the set of clearing vectors is non-empty.

In order to prove that $p^*(c)$ is the \emph{maximal} clearing vector, consider any other clearing vector $\tilde p=T_c(\tilde p)$. Since $p^0=\bar p\geq\tilde p$,
one has $p^1=T_c(p^0)\geq T_c(\tilde p)=\tilde p$ and, using induction, $p^k\geq\tilde p$ for each $k$. Passing to the limit as $k\to\infty$, one has $p^*(c)\geq\tilde p$.

Symmetrically, one can notice that the sequence of vectors
\begin{gather*}
  \tilde p^0(c)=0,\;\; \tilde p^k(c)=T_c(\tilde p^{k-1}(c))\in [0,\bar p],\quad \forall k\geq 1.
\end{gather*}
is non-decreasing $\tilde p^k\leq\tilde p^{k+1}$, because $\tilde p^0\leq\tilde p^1$. Similar to the previous argument, it can be proved that the limit $p_*(c)\doteq\lim_{k\to\infty}\tilde p^k$ is the minimal clearing vector.
\est

\section{Proof of Proposition~\ref{prop:lp-feasible}}\label{sed:proof:prop:lp-feasible}

Consider the maximal clearing vector $p^*(c)$. Notice first that if $c+A^{\top}p^*(c)\geq 0$, then the constraints~\eqref{eq_clearingopt_LP} are feasible
and satisfied, in fact, by $p=p^*(c)$ and by any other clearing vector $p$ such that $c+A^{\top}p\geq 0$. Indeed, every such vector is a solution to~\eqref{eq:clearing-2}, which means that $0\leq p\leq\bar p$ and $p\leq [c+A^{\top}p]^+=c+A^{\top}p$.
Since the polyhedron defined by constraints~\eqref{eq_clearingopt_LP} is compact, the optimization problem~\eqref{eq_clearingopt_LP} has a solution. Retracing the proof of Lemma~4 in~\cite{Eisenberg2001}, one shows that each minimizer $p^*$ has to be a clearing vector, and thus $p^*\leq p^*(c)$. Hence, $p^*=p^*(c)$, otherwise, $p^*(c)$ would provide a smaller value of the loss function, which is strictly decreasing in $p$. We have proved that if $c+A^{\top}p^*(c)\geq 0$, then the LP~\eqref{eq_clearingopt_LP} is feasible and admits the unique minimizer $p^*=p^*(c)$.

By noticing that $d^*=c+A^{\top}p^*(c)$ coincides with the vector~\eqref{eq.vector-d} of residual values, corresponding to clearing vector $p^*(c)$, one proves that
all debts to the external sector are paid, if one applies the clearing vector $p^*(c)$.

On the other hand, if the constraints in~\eqref{eq_clearingopt_LP} are infeasible, the maximal clearing vector $p^*(c)$ fails to satisfy them. Hence,
$c_i+(A^{\top}p^*(c))_i<0$ for some bank $i\in\calV$. Then, for an arbitrary clearing vector
$p$ one has $p\leq p^*(c)$, and thus $c_i+(A^{\top}p)_i<0$. This means that bank $i$ is insolvent upon paying its external debts. %This finishes the proof.
\est

\begin{remark}
Evidently, the first statement of Proposition~\ref{prop:lp-feasible} retains its validity if the loss function~\eqref{eq.loss-function} in~\eqref{eq_clearingopt_LP} is replaced by any continuous decreasing function $\tilde{L}:[0,\bar p] \to \mathbb{R}$, because Lemma~4 in~\cite{Eisenberg2001} does not rely on linearity of
the cost function.
\end{remark}

\section{Proof of Proposition~\ref{prop:eta-concave}}\label{sed:proof:prop:eta-concave}

Consider the set $\mathcal{M}=\{(c,p):c\in\mathbb{R}^m,\,p\in[0,\bar p],\,c+A^{\top}p\geq 0\}$. The mapping $\mathfrak{T}(c,p)=T_c(p)$ introduced in the proof of Proposition~\ref{prop:existence} is, as has been proved, non-decreasing (in two arguments) and is element-wise \emph{concave}, because
the function
\[
(c,p)\mapsto\min(\bar p_i,[c_i+(A^{\top}p)_i]^+)=\min(\bar p_i,c_i+(A^{\top}p)_i)
\]
is concave for each $i$ as the minimum of two concave functions.

By definition, $(c,p^*(c))\in\mathcal{M}$ whenever $c\in\calC_{feas}$. In this situation, for every $p\geq p^*(c)$, one has $(c,p)\in\mathcal{M}$. In particular, the sequence $p^k(c)$ from~\eqref{eq.iter-max-clearing} satisfies the condition
$(c,p^k(c))\in\mathcal{M}$ for all $k$ and all $c\in\calC_{feas}$.

We now introduce the mapping $\mathfrak{T}:\mathcal{M}\to[0,\bar p]$ defined as
$\mathfrak{T}(c,p)\doteq T_c(p)$. Obviously, $\mathfrak{T}$ is non-decreasing, that is, $c\leq c',p\leq p'$ entails that $\mathfrak{T}(c,p)\leq \mathfrak{T}(c',p')$.
Recalling that a composition $f\circ g$ of two non-decreasing concave function $f,g$ is non-decreasing and concave and using induction on $k$,
one easily proves now that
\[
p^k(c)=\mathfrak{T}(c,p^{k-1}(c))
\]
is a non-decreasing element-wise concave function of $c\in\calC_{feas}$ for each $k$. Hence, $p^*(c)=\lim_{k\to\infty}p^k(c)$ is a non-decreasing and element-wise concave function of $c$. The remaining statements about $\eta^*$ are now straightforward.\est
\color{black}

\section{Proof of Theorem~\ref{prop:wc1}}
\label{sed:proof:prop:wc1}
Consider the primal problem \eqref{eq_clearingopt_LP} and let $\alpha\geq 0$, $\beta \geq 0$, $\lam\geq 0$ be $n$-vectors of Lagrange multipliers. The  Lagrangian of this problem is
\[
\calL = \one\tran (\bar p - p)   -\alpha\tran p
+\beta\tran (p-\bar p)  +\lam\tran (p - \bar c - S\delta - A\tran p )
\]
and the corresponding dual function is
\[
g(\alpha,\beta,\lam) = \min_{ p}\;\calL =  \min_{ p}\;
(\beta +\lam - \one -\alpha -A\lam)\tran p + (\one -\beta)\tran \bar p -\lam\tran (\bar c + S\delta)
\]
Now, $g(\lam) = -\infty$ whenever the coefficient of $p$ is nonzero, and it is equal to
$(\one -\beta)\tran \bar p -\lam\tran (\bar c + S\delta)$ otherwise.
The dual optimization problem is therefore
\beas
\max_{\alpha,\beta,\lam\geq 0}&  (\one -\beta)\tran \bar p -\lam\tran (\bar c + S\delta) \\
\mbox{s.t.: } & \beta +\lam - \one -\alpha -A\lam=0,
\eeas
or, equivalently, eliminating $\alpha \geq 0$,
\beas
\max_{\beta,\lam\geq 0}&  (\one -\beta)\tran \bar p -\lam\tran (\bar c + S\delta) \\
\mbox{s.t.: } &\beta - \one +(I-A)\lam \geq 0.
\eeas
This dual problem is always strictly feasible, since  for any $\lam\geq 0$ one can always find a sufficiently large $\beta\geq 0$ such that the constraint is satisfied with strict inequality.
In such case,  strong duality holds (see, e.g., Section 5.2.4 of \cite{Boyd2004}), hence we have that the optimal objective of the above problem is equal to $\eta^*(\delta)$. Therefore, problem (\ref{eq_clearingopt_LP3}) can be formulated equivalently as
 \beas
\eta\ped{wc} = \max_{\|\delta\|\leq \epsilon }\max_{\beta,\lam\geq 0} &
\;(\one -\beta)\tran \bar p -\lam\tran (\bar c + S\delta)   \label{eq_clearingopt_LP4}\\
 \mbox{s.t.: } & \beta - \one +(I -A)\lam\geq 0. \nonumber
 \eeas
 But
 \beas
 \lefteqn{
 \max_{\|\delta\| \leq \epsilon }  \;
(\one -\beta)\tran \bar p -\lam\tran (\bar c + S\delta)  } \\
&& \rule{2cm}{0cm} =
 (\one -\beta)\tran \bar p -\lam\tran \bar c   +  \max_{\|\delta\| \leq \epsilon }   -\lam\tran S\delta \\
 && \rule{2cm}{0cm} = (\one -\beta)\tran \bar p -\lam\tran \bar c   -  \min_{\|\delta\| \leq \epsilon }   \lam\tran S\delta \\
 &&  \rule{2cm}{0cm}=
  (\one -\beta)\tran \bar p -\lam\tran \bar c   +
 \epsilon \|S\tran \lam\|_* ,
 \eeas
 where $\|\cdot\|_*$ denotes the dual norm to $\|\cdot\|$.
 The problem now becomes
 \beas
\eta\ped{wc} = \max_{\beta,\lam\geq 0}  & \; (\one -\beta)\tran \bar p -\lam\tran \bar c   +
 \epsilon \|S\tran \lam\|_*
 \label{eq_clearingopt_LP5}\\
 \mbox{s.t.: } & \beta - \one +\lam -A\lam \geq 0\nonumber
 \eeas
 as claimed.
 \est

\begin{remark}\label{rem.linear-cost}
  As follows from Proposition~\ref{prop:lp-feasible}, the cost function $\one\tran(\bar p-p)$ in~\eqref{eq_clearingopt_LP} and~\eqref{eq_clearingopt_LP3} can be replaced by an arbitrary linear cost $\tilde L(p)=\nu-\varkappa^{\top}p$, where $\varkappa>0$ is a positive vector (making $\tilde L$ decreasing) and $\nu\in\mathbb{R}$.
  Retracing the duality argument presented above, the worst-case loss value is found as
  \beas
\max_{\beta,\lam\geq 0}  & \; (\nu -\beta\tran \bar p -\lam\tran \bar c   +
 \epsilon \|S\tran \lam\|_*)
 \label{eq_clearingopt_LP5+}\\
 \mbox{s.t.: } & \beta - \varkappa +\lam -A\lam \geq 0\nonumber
 \eeas
 The result of Theorem~\ref{prop:wc1} corresponds to the special case of $\nu=\one\tran\bar p$ and $\varkappa=\one$.
\end{remark}

 \section{Proof of Theorem~\ref{prop:eps_ub}}
\label{sec:proof:eps_ub}
Consider Problem~\eqref{eq_clearingopt_LP}, with $c= \bar c + S\delta$. For fixed $\epsilon \geq 0$, we have that such problem is feasible for all possible $\delta: \|\delta\|\leq \epsilon$ if and only if
there exist $p\geq 0$, $p\leq \bar p$ such that $\bar c + S\delta + A\tran p \geq p$ for all
$\delta: \|\delta\|\leq \epsilon$. This latter condition is satisfied if and only if
$\min_{\delta: \|\delta\|\leq \epsilon} \bar c_i + \sigma_i\tran \delta + (A\tran p)_i \geq p_i$
for all $i=1,\ldots,n$. Computing the minimum on the left-hand side of this expression we have the equivalent conditions
\[
\bar c_i - \epsilon\| \sigma_i\tran \|_* + (A\tran p)_i \geq p_i
\]
for all $i=1,\ldots,n$. For finding $\epsilon\ped{ub}$ we then find the largest $\epsilon$ such that
the previous robust feasibility condition is satisfied, which implies the statement of Theorem~\ref{prop:eps_ub}. \est

\section{Proof of Proposition~\ref{prop:uniqueness}
%On the uniqueness of the worst-case perturbation scenario
}
\label{sec:appendix:uniqueness}
We consider a problem of the form
\bea
\eta\ped{wc} = \max_{\beta,\lam}  &  (\one -\beta)\tran \bar p -
 c\ped{wc}\tran \lam
 \label{eq_clearingopt_LP7+a}\\
 \mbox{s.t.: } & \beta   - \one +(I-A)\lam \geq 0 \nonumber \\
 & \lam, \beta \geq 0. \nonumber
 \eea
 This corresponds to problem \eqref{eq_clearingopt_LP7++}, by taking $c\ped{wc}=\bar c -\epsilon |S|\one$, or to the $i$th subproblem in \eqref{eq_clearingopt_LP7+}, by taking $c\ped{wc}=\bar c -\epsilon |\zeta_i|$. Introduce Lagrange multipliers $p\geq 0$ relative to the first constraint, $v\geq 0$ relative to $\beta\geq 0$, and $w\geq 0$ relative to $\lam\geq 0$. Then, it is easy to verify that the Lagrangian dual of problem \eqref{eq_clearingopt_LP7+a} is
 \bea
\eta\ped{wc} = \min_{p,v,w}  &  \one\tran( \bar p - p)
 \label{eq_clearingopt_LP7+b}\\
 \mbox{s.t.: } &
 p \geq 0  \nonumber \\
 v= &\bar p - p \geq 0 \nonumber \\
 w= &c\ped{wc}-(I-A)\tran p \geq 0.\nonumber
 \eea
Observe that for $\epsilon \leq \epsilon\ped{ub}$, where $\epsilon\ped{ub}$ is the insolvency resilience margin, the corresponding  worst-case $c$ value used in
\eqref{eq_clearingopt_LP7+b} is such that this LP remains feasible. Therefore, since
\eqref{eq_clearingopt_LP7+b} actually coincides with the LP \eqref{eq_clearingopt_LP}
discussed in Proposition~\ref{prop:lp-feasible} with $c=c\ped{wc}$, we have that
\eqref{eq_clearingopt_LP7+b} has a unique optimal solution $p^*\ped{wc}$, which represents the clearing vector under the worst-case asset price scenario, in the considered norm. Problem \eqref{eq_clearingopt_LP7+a} and \eqref{eq_clearingopt_LP7+b} are dual to each other, and strong duality holds.

We are here concerned with the uniqueness of the optimal
 solution of  problem \eqref{eq_clearingopt_LP7+a}.
We next  provide a condition for uniqueness which can be tested a posteriori, i.e., after an optimal
solution  has been computed. Let then such an optimal solution $(\beta^*,\lam^*)$ for problem \eqref{eq_clearingopt_LP7+a} be given, and let $(p^*, v^*,w^*)$ be the (unique) corresponding optimal solution
of the dual problem \eqref{eq_clearingopt_LP7+b}. Define the block matrix $F$ and vector $g$ as
\[
F \doteq \left[ \ba{cc} I & I-A \\ I & 0 \\ 0 & I \ea\right], \quad
g \doteq \left[ \ba{c} \one \\ 0 \\ 0 \ea\right], \quad
%h \doteq \left[ \ba{c} -\bar p \\ -c\ped{wc} \ea\right],
\]
so that the constraints in problem \eqref{eq_clearingopt_LP7+a} are rewritten in compact notation as
$F\left[\ba{c}\beta \\\lam \ea\right] \geq g$.
Let $Z$ denote the set of row indices in $F$ such that $F_i\tran \left[\ba{c}\beta^* \\\lam^* \ea\right]= g_i$, where
$F_i\tran$ denotes the $i$th row of $F$, and let
$Z = U\cup L$, where $U$ is the set of indices in $i\in Z$ such that $(p^*, v^*,w^*)_i > 0$,
and $L$ is the set of indices in $i\in Z$ such that $(p^*, v^*,w^*)_i = 0$.
Let $F_U$, $F_L$, $F_Z$  denote  the submatrices obtained from $F$ by selecting the rows of indices in $U$, $L$
and $Z$, respectively. Consider the  LP
\bea
\iota^* = \max_{\beta,\lam}  &  \one \tran F_L \left[\ba{c}\beta \\\lam \ea\right]
\label{eq_uniqcheck} \\
 \mbox{s.t.: } & F_U \left[\ba{c}\beta \\\lam \ea\right]  = 0 \nonumber \\
 & F_L \left[\ba{c}\beta \\\lam \ea\right] \geq 0.\nonumber
 \eea

 \begin{proposition}
 \label{prop_unique}
Given an optimal solution $(\beta^*,\lam^*)$ for problem \eqref{eq_clearingopt_LP7+a}, construct matrices
$F_U$, $F_L$, $F_Z$ and compute the optimal value $\iota^*$ of problem \eqref{eq_uniqcheck}.
The following two conditions are equivalent:
\begin{enumerate}
\item  $F_Z$ has full-column rank and $\iota^*=0$;
\item  $(\beta^*,\lam^*)$ is the unique optimal solution of problem \eqref{eq_clearingopt_LP7+a}.
\end{enumerate}
\end{proposition}
This proposition follows from direct application of Theorem~2 and Remark~2 of \cite{Mangasarian:979}.
We can now derive the following criteria for the uniqueness of the worst-case perturbation scenario that leads to the worst-case loss discussed in Section~\ref{sec:worst-case}.

\section{Core-periphery random graph construction}
\label{sec:appendix:rndgraphconstr}

The structure of the core–periphery graph used in the numerical example of Section~\ref{sec:numeric} follows that of the core–periphery liability network presented in~\cite[Section~3]{hu2024robust}, which is based on bank holding companies (BHCs) that were required to file Federal Reserve Form Y-9C and met specific criteria in 2019. The graph has $n=353$ nodes, where $n_c=29$ nodes form the core of the graph whereas the remaining $n_p=324$ nodes are periphery nodes. Following \cite{hu2024robust}, we make a simplifying assumption that all nodes in the core of the network are connected to every other node, both in the core and in the periphery. Instead, nodes in the periphery
are connected to every node in the core but no other
node in the periphery.

The nominal interbank liabilities $\bar{p}_{ij}$ for every edge $(i,j)$ of the core-periphery graph are then found by sampling from a uniform distribution $\bar{p}_{ij}\sim\mathcal{U}(0,p_{ij}^{max})$, where the upper bound $p_{ij}^{max}$ is defined as
\[
p_{ij}^{max}=\begin{cases}\frac{2p_{cc}}{n_c(n_c-1)}\mathrm{TOT}&\mbox{if }i\mbox{ and } j\mbox{ are core nodes},
\\\frac{2p_{cp}}{n_c n_p}\mathrm{TOT}&\mbox{if }i\mbox{ is a core node and } j\mbox{ is a periphery node},
\\\frac{2p_{pc}}{n_c n_p}\mathrm{TOT}&\mbox{if }i\mbox{ is a periphery node and } j\mbox{ is a core node},\end{cases}
\]
with $\mathrm{TOT}=10000$ representing the aggregate value of all liabilities across the entire network, $p_{cc}=0.789$, $p_{cp}=0.093$, and $p_{pc}=0.118$. This setup allows us to replicate the same proportions of aggregate liabilities between core and peripheral nodes as reported in~\cite[Table 1]{hu2024robust}.

The matrix $S$ of asset shares is a random $50 \times 5$ sparse matrix with a density of 0.4. The nonzero entries $s_{ij}$ are sampled from a uniform distribution, $s_{ij} \sim \mathcal{U}(0, s^{\mathrm{max}})$, with $s^{\mathrm{max}} = 1000$. The external outflows $c_i^-$ are also drawn from a uniform distribution, $c_i^- \sim \mathcal{U}(0, c^{-\mathrm{max}})$, with $c^{-\mathrm{max}} = 1000$. To ensure a default-free performance in the nominal conditions (Assumption~\ref{ass:nodefnom}), the external inflows $c_i^+$ are configured such that, in the nominal operations, the total (nominal) outflow in~\eqref{eq.in-out-nominal} does not exceed the total (nominal) inflow: $\bar\phi_i^{\mathrm{in}}-\bar\phi_i^{\mathrm{out}}\ge 0$. We guarantee this by choosing
\[
c^+_i=[c^-_i+\bar p_i-\sum\nolimits_{j}\bar p_{ji}]^+ + n_i,
\]
%at each node, the total inflow equals or exceeds the total outflow.\[
%c^+_i=\begin{cases} 0 &\mbox{if }\phi_i^{\mathrm{in}}-\phi_i^{\mathrm{out}}\ge 0,\\
%\phi_i^{\mathrm{out}}-\phi_i^{\mathrm{in}}+n_i&\mbox{if }\phi_i^{\mathrm{in}}-\phi_i^{\mathrm{out}}< 0,\end{cases}
%\]
where $n_i\sim\mathcal{U}(0,100)$ is a random variable.
\color{black}

\bibliography{financial}

\end{document}